We, 31 May 1995   14:09:45                    MIDAS version: 94MAY

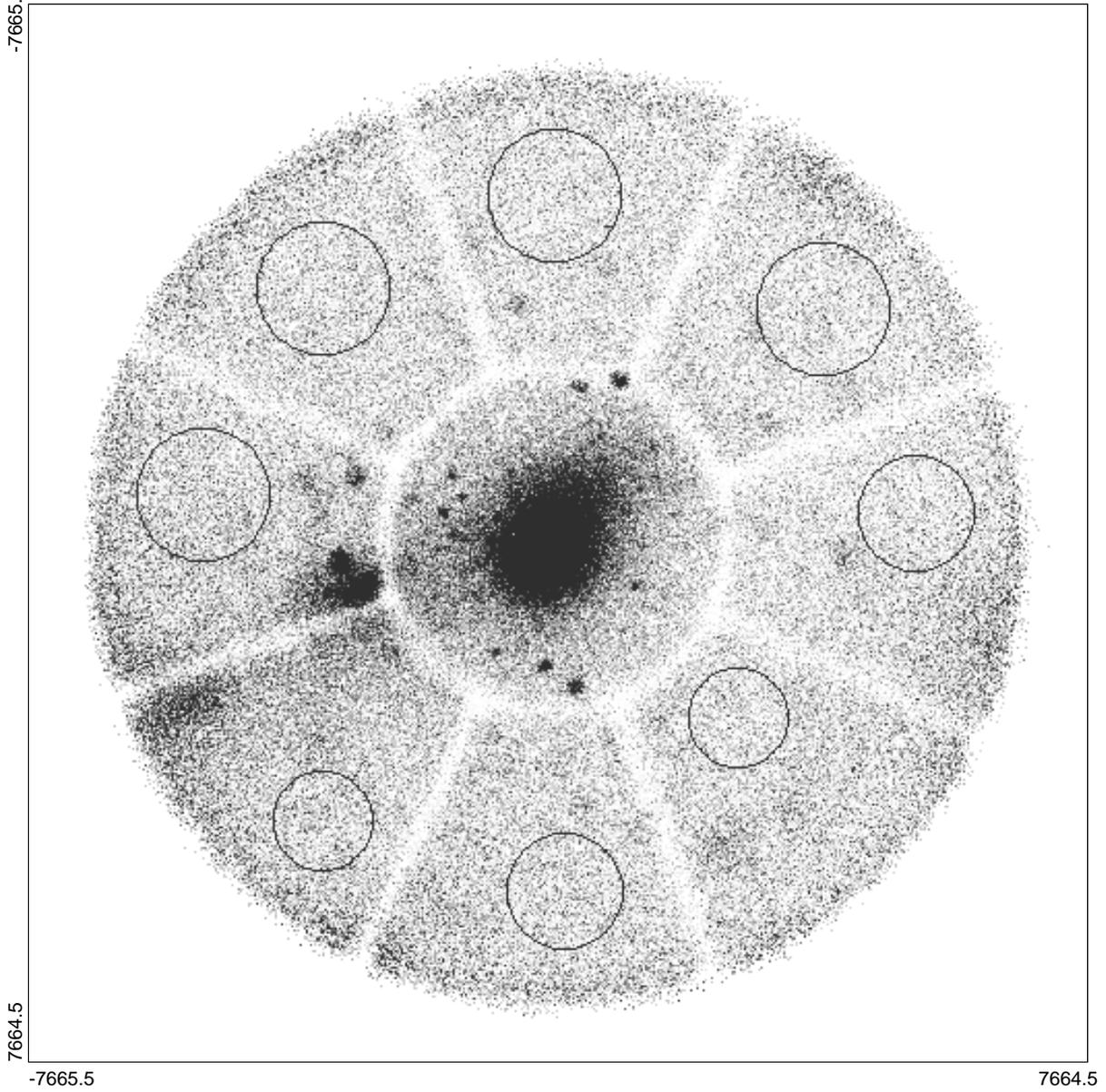

-7665.5                                                         7664.5

Frame       : za3558
Identifier  :
ITT-table   : ramp.itt
Coordinates : -7665.5, 7664.5 : 7664.5, -7665.5
Pixels      : 1, 1 : 512, 512
Cut values  : 1, 9
User        : bardelli

ESO-MIDAS(94MAY); We, 31 May 1995  14:09:45

**Table 1:** Counts in the background circles.

| # | Counts/pixel |
|---|---|
| 1 | $1.965 \pm 0.031$ |
| 2 | $1.941 \pm 0.031$ |
| 3 | $1.925 \pm 0.036$ |
| 4 | $2.086 \pm 0.041$ |
| 5 | $2.247 \pm 0.039$ |
| 6 | $2.474 \pm 0.048$ |
| 7 | $2.528 \pm 0.036$ |
| 8 | $2.141 \pm 0.033$ |

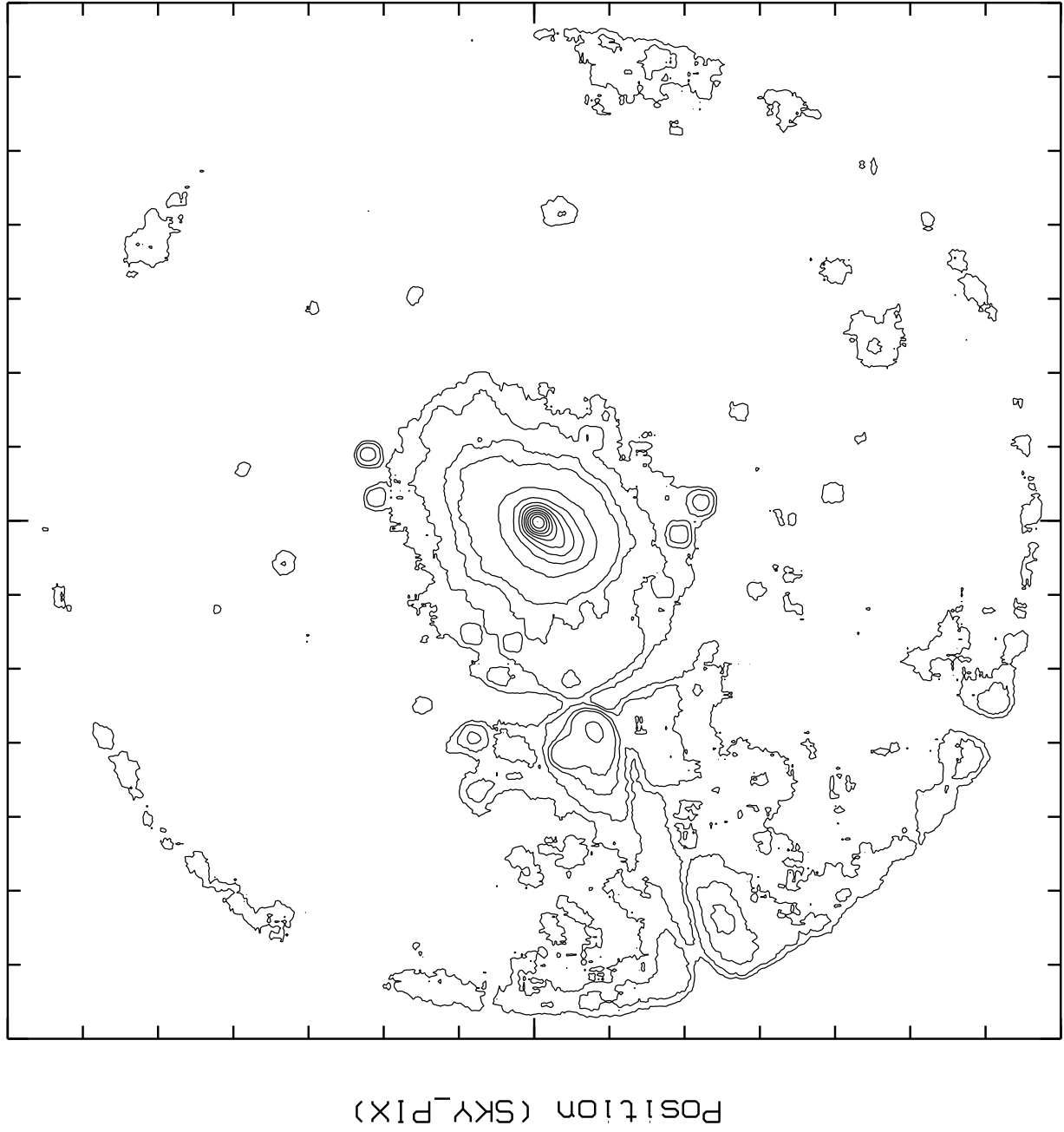

**Table 2:** Results of the bidimensional fit.

| | | |
|---|---|---|
| $x_{cl} = 13^h 28^m 00.37^s$ | $y_{cl} = -31°30'25.1''$ | $I_{cl} = 64.58\ cts/pix$ |
| $x_g = 13^h 27^m 55.21^s$ | $y_g = -31°29'47.6''$ | $I_g = 123.38\ cts/pix$ |
| $\beta = 0.611$ | $\theta = 128°.3$ | $bck = 2.014\ cts/pix$ |
| $R_1 = 14.20\ pix$ | $= 3.55\ arcmin$ | $= 0.148\ Mpc$ |
| $R_2 = 19.58\ pix$ | $= 4.90\ arcmin$ | $= 0.204\ Mpc$ |
| $\sigma = 2.47\ pix$ | $= 0.62\ arcmin$ | $= 0.026\ Mpc$ |

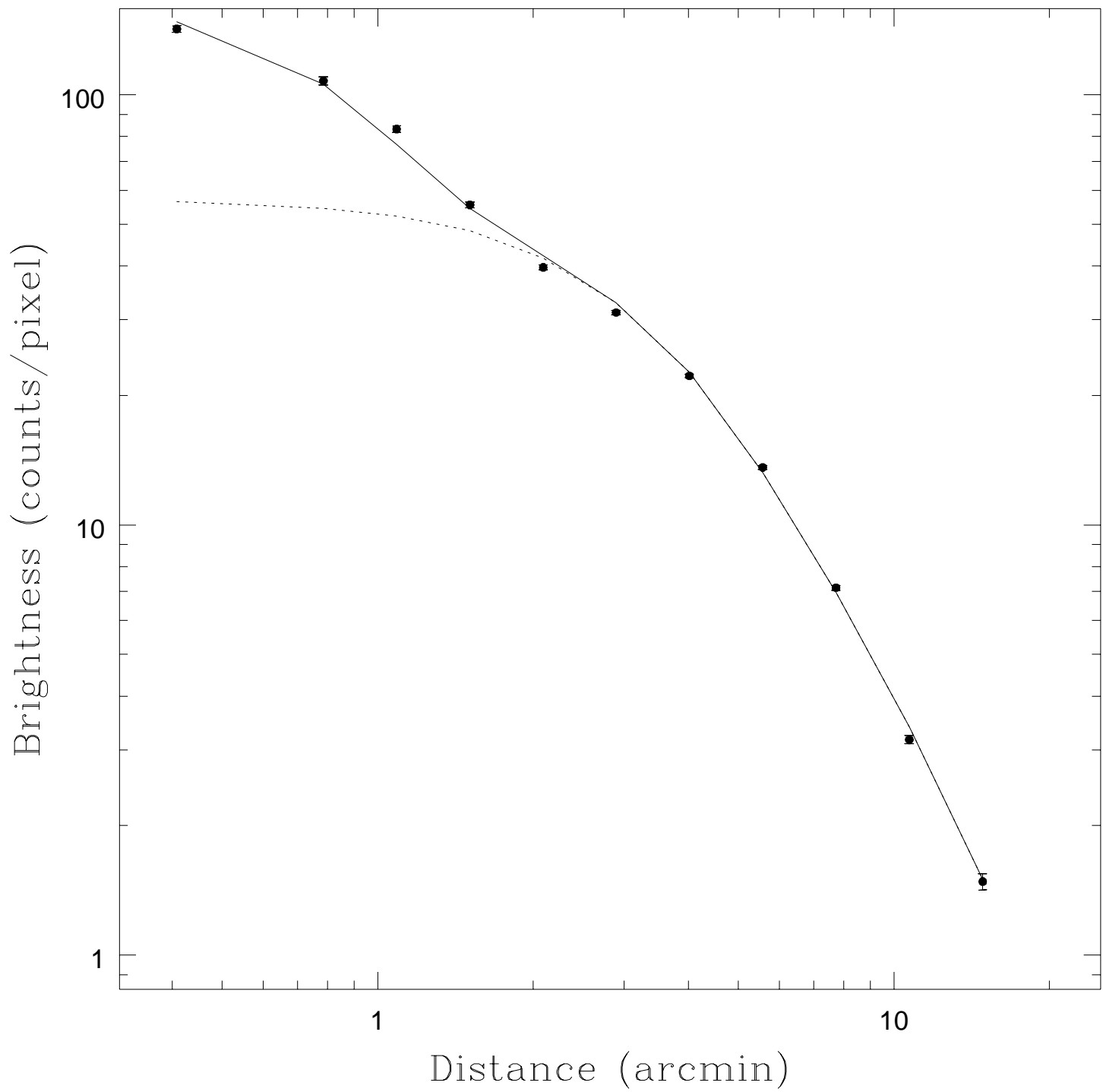

Table 3: Other sources in the map

| # | $\alpha$(2000) h m s | $\delta$(2000) o ′ ″ | offset arcmin | counts | flux[0.5–2.0]keV $10^{-14} erg\ cm^{-2}\ s^{-1}$ |
|---|---|---|---|---|---|
| 1  | 13 25 15.1 | −31 31 54 | 34 | 275   | 11.8  |
| 2  | 13 26 19.5 | −32 07 49 | 43 | 799   | 34.2  |
| 3  | 13 27 11.3 | −31 35 48 | 11 | 150   | 6.4   |
| 4  | 13 27 20.7 | −31 10 46 | 20 | 162   | 6.9   |
| 5  | 13 27 42.8 | −31 11 48 | 18 | 176   | 7.5   |
| 6  | 13 27 45.0 | −31 47 53 | 18 | 339   | 14.5  |
| 7  | 13 27 00.0 | −31 22 22 | 14 | 68    | 2.9   |
| 8  | 13 27 02.0 | −31 18 51 | 14 | 103   | 4.4   |
| 9  | 13 28 02.3 | −31 45 26 | 16 | 335   | 14.3  |
| 10 | 13 28 11.9 | −31 34 50 | 7  | 165   | 7.1   |
| 11 | 13 28 30.5 | −31 43 36 | 16 | 75    | 3.2   |
| 12 | 13 28 34.3 | −31 30 06 | 9  | 132   | 5.6   |
| 13 | 13 28 48.9 | −31 25 11 | 12 | 96    | 4.1   |
| 14 | 13 28 49.7 | −31 30 00 | 12 | 149   | 6.3   |
| 15 | 13 28 54.7 | −31 22 23 | 15 | 132   | 5.6   |
| 16 | 13 28 16.0 | −31 01 42 | 29 | 202   | 8.6   |
| 17 | 13 28 59.7 | −31 26 59 | 14 | 253   | 10.8  |
| 18 | 13 29 47.0 | −31 36 29 | 26 | 15916 | 681.0 |
| 19 | 13 29 49.5 | −31 22 36 | 26 | 491   | 21.0  |
| 20 | 13 29 59.0 | −31 32 42 | 27 | 1617  | 69.1  |
| 21 | 13 30 16.9 | −31 23 27 | 31 | 379   | 16.2  |
| 22 | 13 31 28.0 | −31 49 50 | 49 | 6674  | 285.0 |

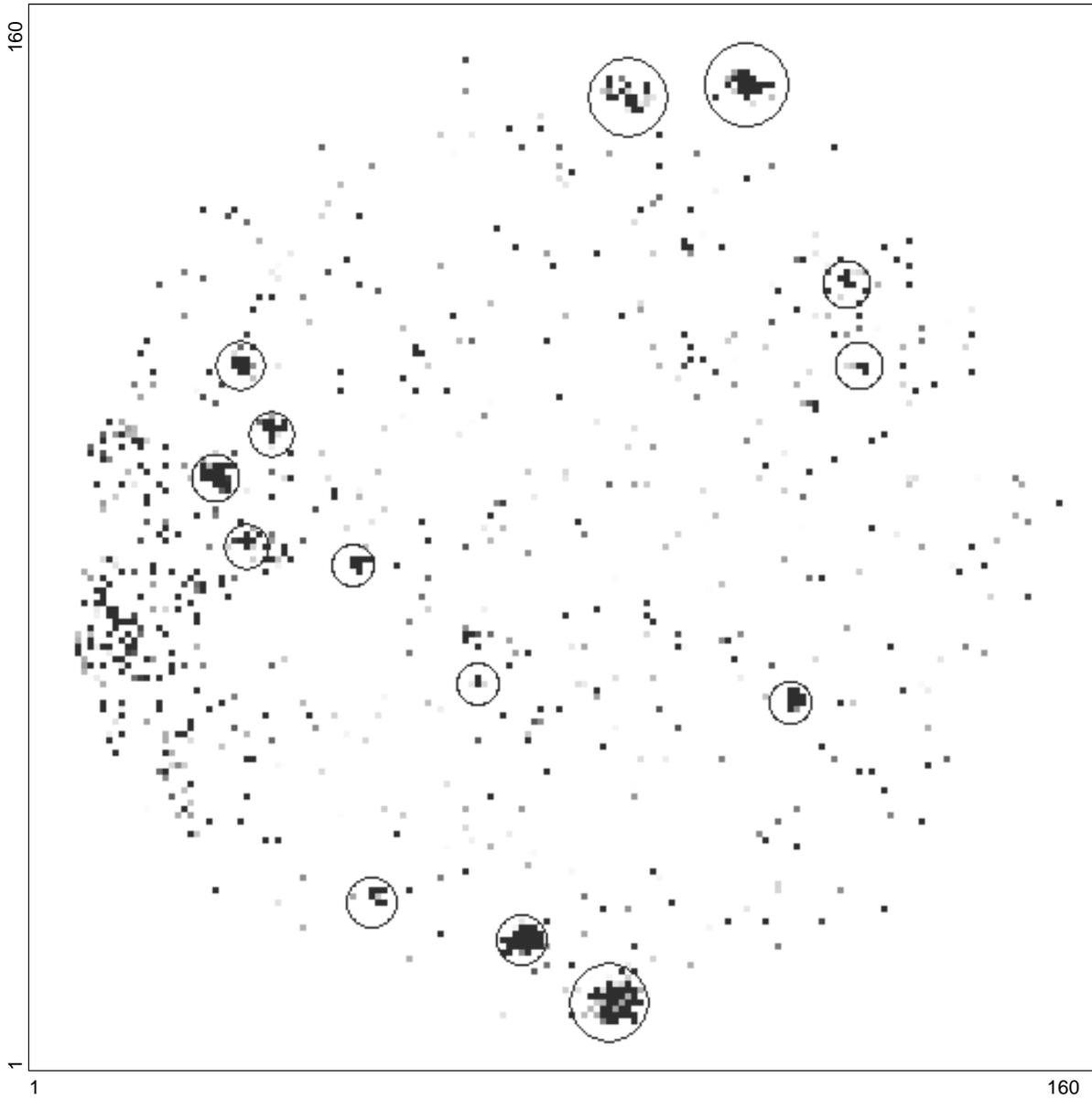

Frame       : reschi
Identifier  :
ITT-table   : ramp.itt
Coordinates : 1, 1 : 160, 160
Pixels      : 1, 1 : 512, 512
Cut values  : 5, 8
User        : comastri

**Table 4a:** Summary of best–fit spectral parameters ($drmpspc_{06}$)

| radius (arcmin) | counts | temperature (keV) | abundance (cosmic=1) | $\chi^2$ |
|---|---|---|---|---|
| cD galaxy | 2824 | $1.89^{+0.86}_{-0.37}$ | $0.61^{+0.38}_{-0.22}$ | 0.60 |
| 0.0 – 2.0 [a] | 9427 | $3.34^{+0.77}_{-0.61}$ | $0.42^{+0.24}_{-0.26}$ | 0.90 |
| 2.0 – 3.5 [a] | 13627 | $3.38^{+0.60}_{-0.48}$ | $0.24^{+0.19}_{-0.23}$ | 0.88 |
| 3.5 – 5.0 | 13858 | $3.54^{+0.66}_{-0.54}$ | $0.39^{+0.22}_{-0.24}$ | 0.80 |
| 5.0 – 8.0 | 19467 | $2.97^{+0.41}_{-0.37}$ | $0.26^{+0.14}_{-0.15}$ | 1.33 |
| 8.0 – 13.0 | 16236 | $3.02^{+0.61}_{-0.47}$ | $0.31^{+0.20}_{-0.20}$ | 1.34 |
| 13.0 – 18.4 | 9972 | $2.23^{+0.63}_{-0.42}$ | $0.21^{+0.21}_{-0.20}$ | 0.81 |
| 13.0 – 18.4 [b] | 6936 | $3.06^{+1.64}_{-0.91}$ | 0.32 fixed | 0.91 |
| $SC\ 1327-312$ | 3512 | $3.85^{+2.40}_{-1.32}$ | $1.06^{+1.46}_{-0.70}$ | 0.51 |
| $HD\ 117310$ | 2139 | $1.10^{+0.11}_{-0.11}$ | $0.19^{+0.09}_{-0.06}$ | 0.97 |

[a] without a sector including the dominant galaxy

[b] without a sector including the poor cluster $SC\ 1327-312$

**Table 4b:** Summary of best–fit spectral parameters ($drmpspc_{36}$)

| radius (arcmin) | counts | temperature (keV) | abundance (cosmic=1) | $\chi^2$ |
|---|---|---|---|---|
| cD galaxy | 2824 | $2.01^{+1.05}_{-0.43}$ | $0.76^{+0.53}_{-0.29}$ | 0.60 |
| 0.0 − 2.0 [a] | 9427 | $3.72^{+0.99}_{-0.71}$ | $0.60^{+0.31}_{-0.32}$ | 0.93 |
| 2.0 − 3.5 [a] | 13627 | $3.79^{+0.78}_{-0.61}$ | $0.39^{+0.24}_{-0.28}$ | 0.92 |
| 3.5 − 5.0 | 13858 | $3.90^{+0.82}_{-0.63}$ | $0.56^{+0.29}_{-0.29}$ | 0.72 |
| 5.0 − 8.0 | 19467 | $3.24^{+0.52}_{-0.39}$ | $0.39^{+0.18}_{-0.18}$ | 1.20 |
| 8.0 − 13.0 | 16236 | $3.42^{+0.70}_{-0.57}$ | $0.51^{+0.24}_{-0.26}$ | 1.36 |
| 13.0 − 18.4 | 9972 | $2.43^{+0.76}_{-0.49}$ | $0.33^{+0.30}_{-0.24}$ | 0.87 |
| 13.0 − 18.4 [b] | 6936 | $3.15^{+2.00}_{-0.95}$ | 0.49 fixed | 0.97 |
| $SC\ 1327 - 312$ | 3512 | $4.45^{+2.72}_{-1.61}$ | $1.68^{+2.94}_{-0.97}$ | 0.66 |
| $HD\ 117310$ | 2139 | $1.14^{+0.10}_{-0.12}$ | $0.23^{+0.10}_{-0.07}$ | 0.88 |

[a] without a sector including the dominant galaxy

[b] without a sector including the poor cluster $SC\ 1327 - 312$

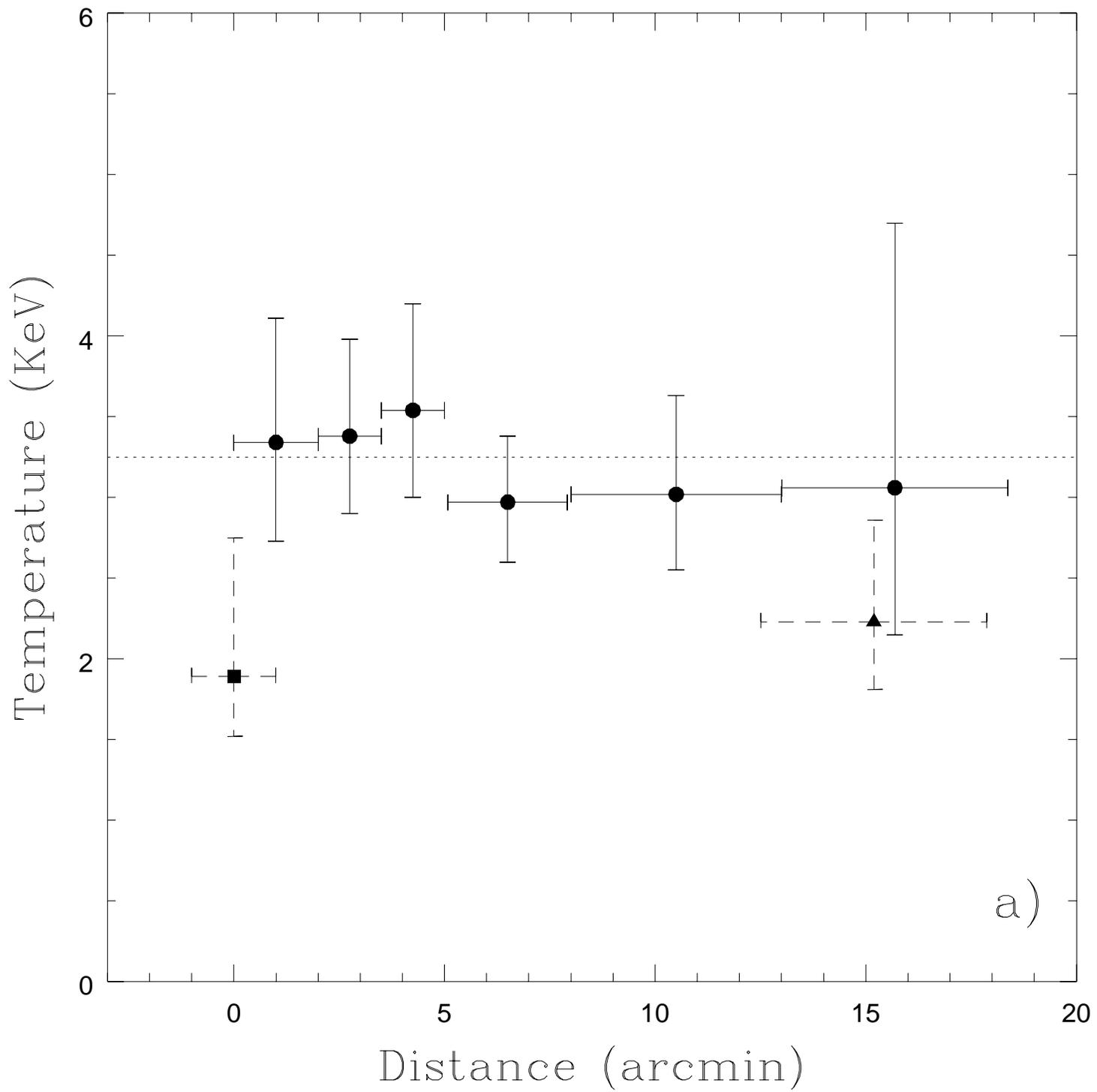

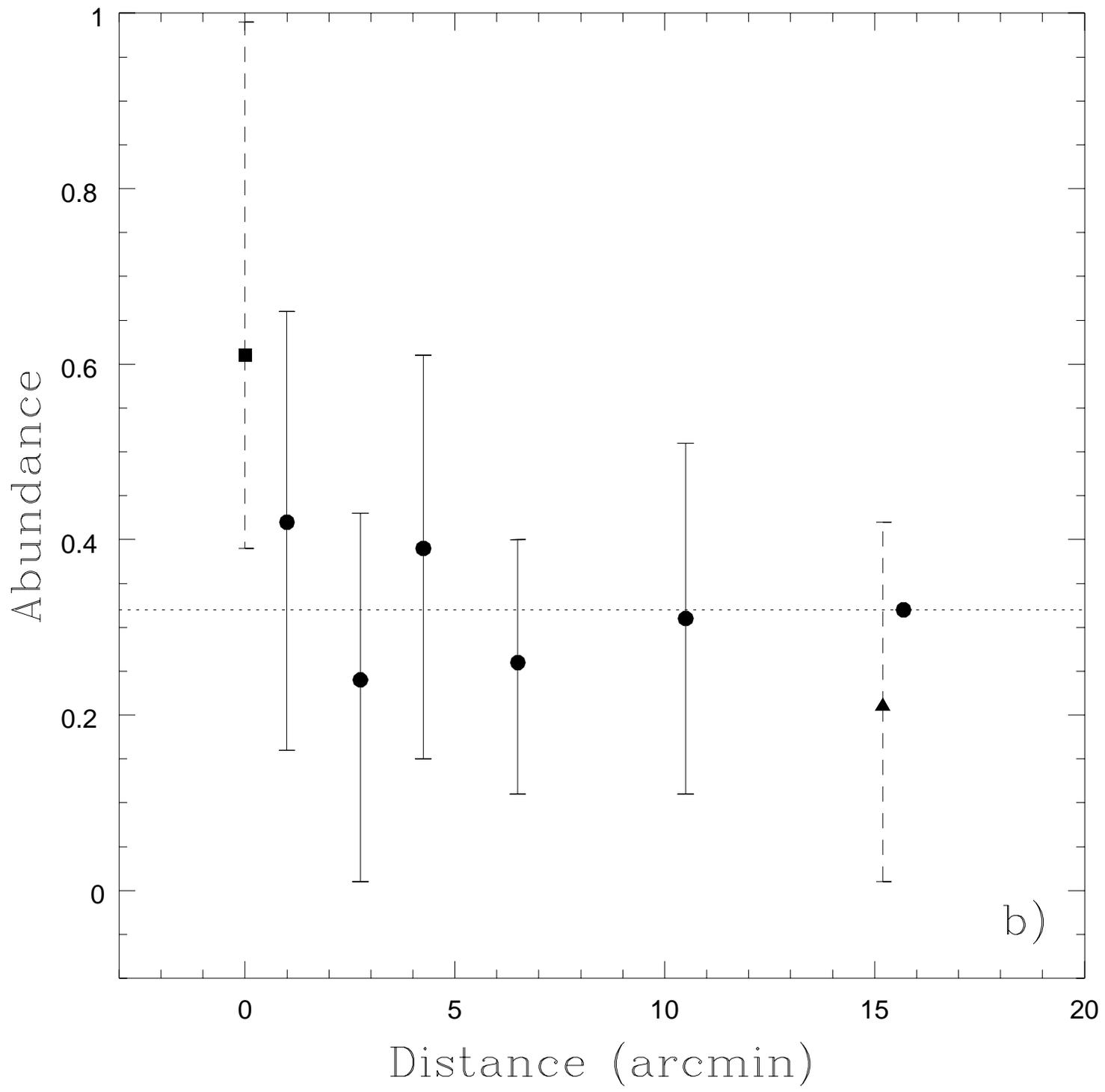

b)

**Table 5:** Mass estimates

| Distance $h^{-1}$ $Mpc$ | Gas Mass $h^{-5/2}$ $10^{13}$ $M_\odot$ | Total Mass $h^{-1}$ $10^{13}$ $M_\odot$ | $M_{gas}/M_{tot}$ $h^{-3/2}$ |
|---|---|---|---|
| 0.50 | 1.5 | 9.4 | 15.7% |
| 0.75 | 2.8 | 15.0 | 18.7% |
| 1.00 | 4.3 | 20.4 | 20.8% |
| 1.25 | 5.8 | 25.8 | 22.4% |
| 1.50 | 7.4 | 31.2 | 23.7% |

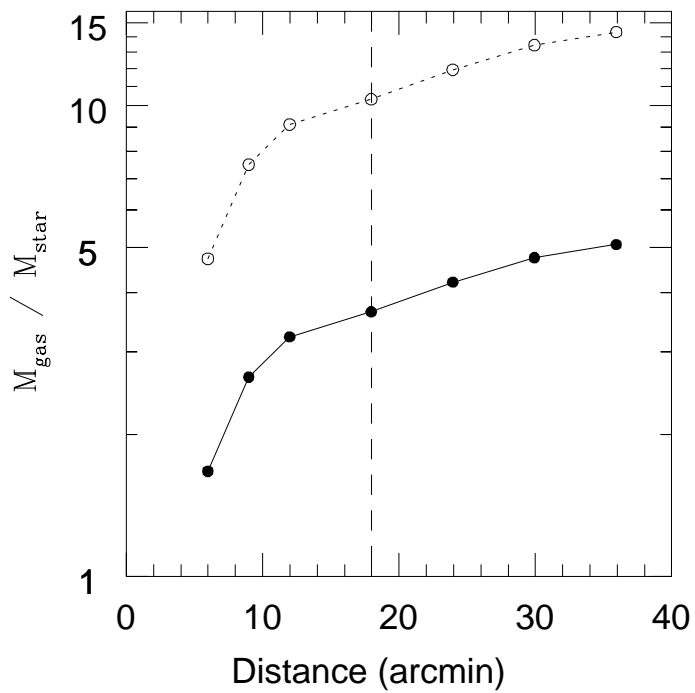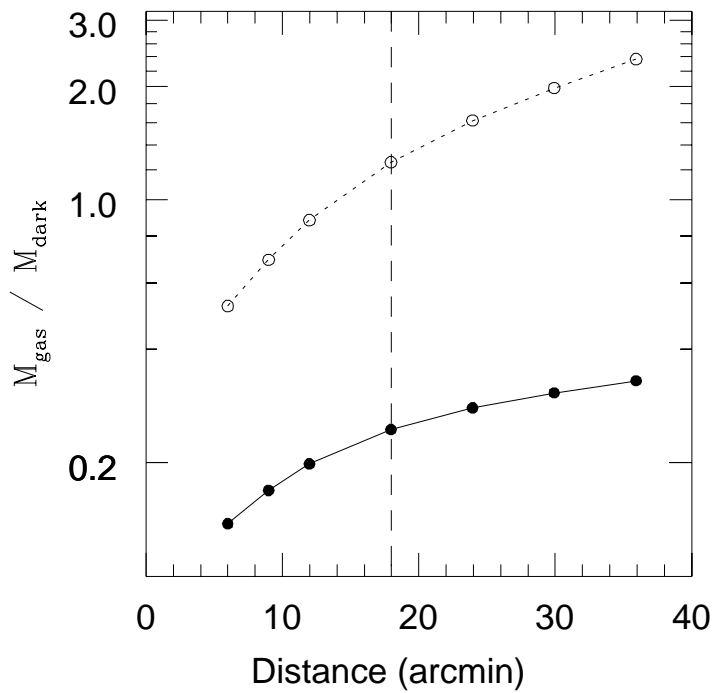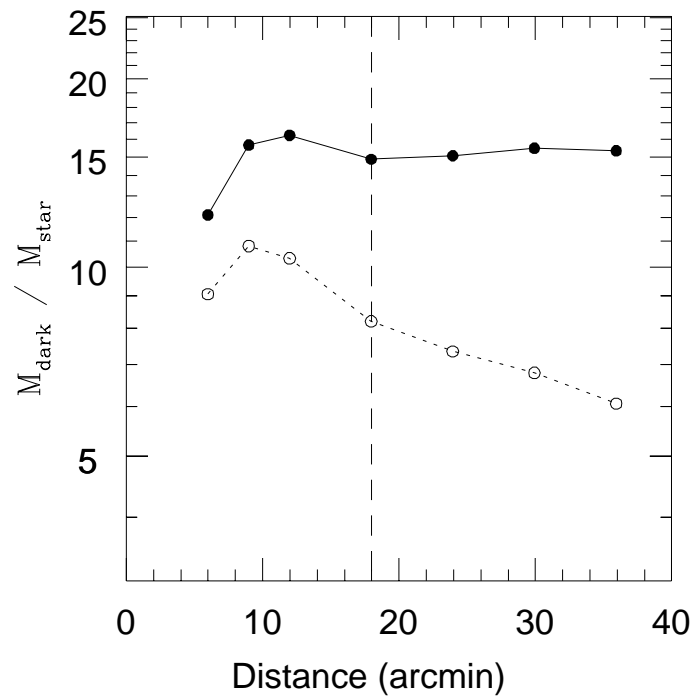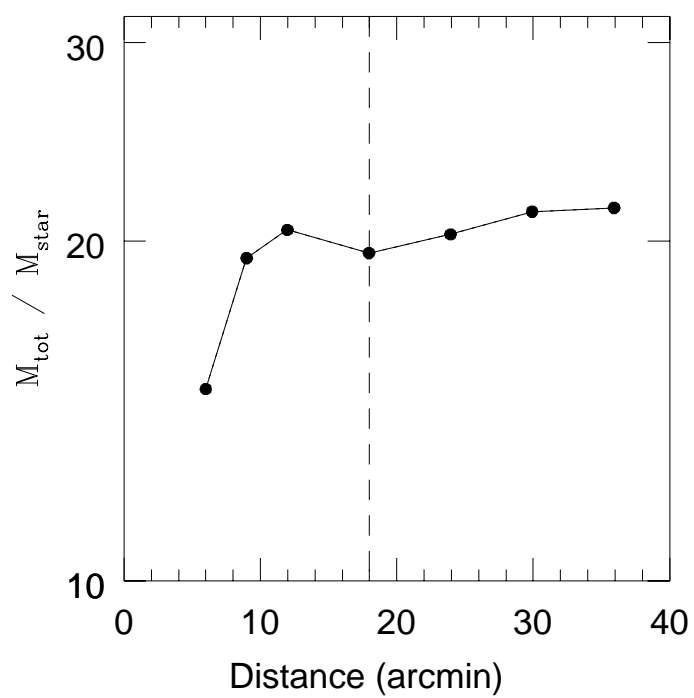

# A STUDY OF THE CORE OF THE SHAPLEY CONCENTRATION: II. ROSAT OBSERVATION OF A3558

*Sandro BARDELLI*

Osservatorio Astronomico, via Zamboni 33, 40126 Bologna, Italy
and
Istituto di Radioastronomia/CNR, via Gobetti 101, 40129 Bologna, Italy.

*Elena ZUCCA*

Istituto di Radioastronomia/CNR, via Gobetti 101, 40129 Bologna, Italy.

*Angela MALIZIA*

Dipartimento di Astronomia, Università di Bologna, via Zamboni 33, 40126 Bologna, Italy.

*Giovanni ZAMORANI*

Osservatorio Astronomico, via Zamboni 33, 40126 Bologna, Italy
and
Istituto di Radioastronomia/CNR, via Gobetti 101, 40129 Bologna, Italy.

*Roberto SCARAMELLA*

Osservatorio Astronomico di Roma, 00040 Monteporzio Catone, Italy.

*Giampaolo VETTOLANI*

Istituto di Radioastronomia/CNR, via Gobetti 101, 40129 Bologna, Italy.

# ABSTRACT

We present the results of a pointed ROSAT PSPC observation of the Abell cluster *A3558*. *A3558* is the only cluster classified as richness class four in the entire ACO catalogue and, together with *A3556* and *A3562*, it is part of an elongated structure defining the core of the Shapley Concentration.

The X-ray surface brightness distribution of the cluster can be fitted as the sum of two components: an elliptical King law, with a major core radius of $\sim 0.20\ h^{-1}\ Mpc$ and an axial ratio $\sim 0.72$, and a Gaussian central source associated with the cD galaxy. The centers of the two X-ray components are separated by 75 $arcsec$, corresponding to $\sim 50\ h^{-1}\ kpc$. A number of additional discrete sources is also present in the field. Although most of them are probably unrelated to the cluster, two of these sources are identified with the poor clusters $SC\ 1327 - 312$ and $SC\ 1329 - 313$, at approximately the same redshift as *A3558*, and two other sources are identified with bright galaxies, which are also part of the core of the Shapley Concentration. In addition, there is also evidence of an enhanced diffuse X-ray emission which connects the outer part of *A3558* with $SC\ 1327 - 312$.

From the spectral analysis, we find that the temperature profile of *A3558* is consistent with an isothermal distribution at $kT \sim 3.25\ keV$, while the cD galaxy has the lower temperature $kT = 1.89\ keV$. For the cluster we derive a luminosity $L_X = 1.1 \times 10^{44}\ h^{-2}\ erg\ s^{-1}$, in the energy range $[0.5 - 2.0]\ keV$, within a radius of $0.8\ h^{-1}\ Mpc$ and a total mass $M_{tot} = 3.1 \times 10^{14}\ h^{-1}\ M_\odot$ within an Abell radius ($1.5\ h^{-1}\ Mpc$). In the same region the gas mass is $\sim 20\%\ h^{-3/2}$ of the total mass and there are evidences for the hot gas to be less centrally concentrated than the dark matter and the stellar component. For $h = 1$ most of the dynamical mass is in dark matter, while for $h = 0.5$ most of the mass is in hot gas. In both cases, the stellar mass appears to be a good tracer of the total mass distribution.

The derived $M/L$ ratio is $\sim 130\ h\ M_\odot/L_\odot$ for radii larger than $\sim 0.4\ h^{-1}\ Mpc$.

Finally, under the hypothesis that the central density spike represents a cooling flow, we derive a mass deposition rate of $\sim 25\ h^{-2}\ M_\odot\ yrs^{-1}$.

The single physical parameters derived for *A3558* show a general agreement with the typical values found for other X-ray emitting clusters, but the correlations between these quantities show some peculiarities, probably due to the complex dynamical state of this cluster.

# 1. INTRODUCTION

From studies of the cluster formation rate as a function of the mean density of the Universe (Richstone et al. 1992, Lacey & Cole 1993, and references therein), it results that the bulk of the cluster formation happens later when the value of $\Omega$ is higher. Therefore the central parts of rich superclusters, which have a local density higher than the mean and so can be considered similar to high $\Omega$ universes (Gunn & Gott 1972), are nearby "laboratories" where it is possible the study of the dynamical phenomena concerning the evolution of massive clusters. Indeed in these environments the high density, combined with the high peculiar velocities, increases the probability of interactions.

The Shapley Concentration is the richest supercluster which appears studying the distribution of the optically selected Abell–ACO clusters of galaxies (Scaramella et al. 1989; Zucca et al. 1993). It is very rich also in X–ray clusters, containing six of the 46 clusters of the all–sky sample of Edge et al. (1990), which is limited to $1.7 \times 10^{-11}$ $erg$ $cm^{-2}$ $s^{-1}$ in the energy band $[2 - 10]$ $keV$ (see also Lahav et al. 1989 and Raychaudhury et al. 1991). Using the relationship between X–ray luminosity and gas mass for clusters, Fabian (1991) has estimated the ratio of the baryon density to the critical density in the 17.5 $h^{-1}$ $Mpc$ (hereafter $h = H_o/100$) core of the Shapley supercluster to be a factor at least three higher than the value predicted by the cosmic nucleosynthesis. Such a high baryon content requires that the supercluster has a total mass overdensity $\rho/\rho_o > 2.4$, the largest yet established on such large scales. A similar excess of baryonic matter with respect to the predictions of the cosmic nucleosynthesis is also seen in rich clusters. White et al. (1993b) have suggested that this excess can not be accounted for by gravitational and dissipative processes during cluster formation and may require either a low–density Universe ($\Omega_o \sim 0.3$) or some non–standard models for the cosmic nucleosynthesis.

Raychaudhury et al. (1991) noted that the fraction of X–ray multiple clusters in this supercluster is greater than 50%, i.e. much higher than the average value in the field ($\sim 10\%$), and that the X–ray luminosity function, which is a factor between 10 and 50 higher than the average, suggests a merging rate between 1.5 and 3 times higher than elsewhere; in other words, interesting dynamical processes, like cluster–cluster and group–cluster mergings, are probably present in this region.

The inner core of the Shapley Concentration is individuated by a remarkable chain formed by the three ACO clusters $A3556$, $A3558$ and $A3562$, at a distance of $\sim 140$ $h^{-1}$ $Mpc$. Using both photometric and spectroscopic data, Bardelli et al. (1994, hereafter Paper I) found that these three clusters are likely to be strongly interacting and form a single elongated structure, almost orthogonal to the line of sight. In this structure, in addition to the main clusters, also a number of minor subcondensations are present. In particular, they noted two subclumps in–between $A3558$ and $A3562$, apparently separated in velocity, whose galaxies show some evidence of different luminosity distributions. These facts suggest a quite complicated dynamical state for this structure: Metcalfe et al. (1994) suggested that there may have already been an encounter between $A3558$ and $A3562$.

The chain is dominated by $A3558$ (also known as Shapley 8), the richest cluster of the ACO catalogue (Abell et al. 1989), the only one classified as richness class 4. It has been studied in the optical band by Metcalfe et al. (1987), Teague et al. (1990) and Metcalfe et al. (1994); in Paper I, using more than 250 redshifts of galaxies belonging to the cluster, we estimated a central velocity dispersion of the galaxies of $\sim 1000\ km/s$. Metcalfe et al. (1987), on the basis of their velocity data for 40 galaxies in the core of the cluster, derived a blue luminosity of $\sim 3.7 \times 10^{11}\ h^{-2}\ L_\odot$ and, using an estimator related to the virial mass, obtained an $M/L$ ratio of the order of $(392 \pm 120)\ h$, within 5 $arcmin$ from the cluster center. Metcalfe et al. (1994) derived the total blue luminosity as a function of the radius, estimating $M/L \sim 560\ h$ within $1.5\ h^{-1}\ Mpc$. Biviano et al. (1993) calculated the virial mass, obtaining $6 \times 10^{14}\ h^{-1}\ M_\odot$ within $0.75\ h^{-1}\ Mpc$.

$A3558$ is dominated by a central giant galaxy, classified as a cD galaxy by Metcalfe et al. (1994), which is suspected to have a significant peculiar velocity with respect to the average velocity of the cluster (Gebhardt & Beers 1991). In Paper I, however, we warned about the possibility that the cluster parameters may be somewhat contaminated by the inclusion in the cluster sample of some galaxies belonging to nearby groups. Indeed, by dividing our large velocity sample in two sub–samples, eastward and westward of the center of $A3558$, we found that the velocity of the dominant galaxy is consistent with the average velocity derived for the eastward sample.

This cluster was already observed in the X–ray band by the ME instrument of EXOSAT (Edge & Stewart 1991a) in the energy range $[2 - 10]\ keV$. Although the cluster center was offset by $\sim 38\ arcmin$ from the field center, the authors could estimate a temperature of $3.7^{+1.0}_{-2.0}\ keV$, a corrected flux of $4.2 \times 10^{-11} erg\ cm^{-2}\ s^{-1}$ and a luminosity of $1.1 \times 10^{44}\ h^{-2}\ erg\ s^{-1}$. Day et al. (1991), using the LAC instrument of the GINGA satellite (energy range $[1.7 - 37.8]\ keV$), found that the spectrum of *A3558* is well fitted by two components at $6.2 \pm 0.3$ and $2\ keV$ respectively, and reported a global luminosity in the $[2 - 10]\ keV$ energy range of $2.2 \times 10^{44}\ h^{-2}\ erg\ s^{-1}$. This measurement, significantly higher than the corresponding EXOSAT value, is probably contaminated by emission from the nearby clusters, not spatially resolved by the LAC instrument. More recently Breen et al. (1994), analysing the data from the EINSTEIN slew survey, found a luminosity within $0.5\ h^{-1}\ Mpc$ of $1.35 \pm 0.35 \times 10^{44}\ h^{-2}\ erg\ s^{-1}$ in the band $[0.5 - 4.5]\ keV$.

In this paper we present the analysis of a ROSAT PSPC observation, pointed on *A3558*. In Section 2 we describe the observation and the main procedures of the data reduction, while in Sections 3 and 4 we present the results obtained from the spatial and spectral analyses, respectively. We then derive estimates of the gas and total masses and their radial distribution (Section 5). Finally in Section 6 we summarize and discuss our results.

## 2. THE OBSERVATION AND THE DATA REDUCTION

The observation was made with the Position Sensitive Proportional Counter (hereafter PSPC) on board of the X-ray astronomy satellite ROSAT (Trümper 1983). The PSPC combines good spatial ($\sim 30\ arcsec$, FWHM) and moderate spectral $[\Delta E/E = 0.43(E/0.93\ keV)^{-0.5}$, FWHM] resolution in the energy range $[0.1 - 2.4]\ keV$. The observation was taken on July 17 and 18, 1991, and no wobbling of the support grid of the PSPC window was present. The nominal pointing center was at $\alpha(2000) = 13^h27^m55^s.2$ and $\delta(2000) = -31°29'24''$, i.e. on the cluster position as given in the ACO catalogue, and the total effective exposure time was $30,213$ seconds.

The reduction was performed using the Extended Scientific Analysis System (hereafter EXSAS) package written by Zimmerman et al. (1992) and implemented in the Midas environment: the main steps of the reduction procedure can be summarized as follows.

First, in order to exclude the energy channels which are either not correctly calibrated or significantly contaminated by particle background, we have restricted our analysis to the energy range $[0.14 - 2.13]\ keV$. Then, following Snowden et al. (1994), we have eliminated all the time intervals in which either the background count rate, estimated in a large annulus outside the ribs, is more than 30% higher than the mean value or the Master Veto rate is higher than $200\ counts\ s^{-1}$. Most of the time intervals eliminated in this way correspond to the start or end of the various single observation intervals. As a consequence of this correction, which excludes from the following analysis all the time intervals with strong particle and solar background, we have eliminated $\sim 22\%$ of the background counts, while reducing the effective exposure time by $\sim 15\%$, from $30,213$ to $25,481$ seconds, and thus increasing the final signal to noise ratio of the observation.

Finally, we checked the after–pulse contamination, counting how many events are followed by another one within a time interval of $0.35\ ms$. We found only 122 such events within the central 20 $arcmin$ and 226 in the outer region. Since this contamination is negligible with respect to the following analysis, we have not applied any further correction for this effect.

After this procedure, the counts were extracted from the event table and corrected for vignetting and dead time, using the standard correction mode of EXSAS. Figure 1 shows the resulting image of the whole PSPC field (2 degrees diameter). This image, with North at the top and East to the left, has been obtained by applying a factor of 30 compression to the event table data, which are listed in detector pixels of $0.5 \times 0.5\ arcsec^2$. The resulting pixel of this image is therefore $15 \times 15\ arcsec^2$. Because of the absence of the wobbling during the observation, all the eight ribs are clearly visible in the figure. The higher than average counts which are seen in the outer $\sim 3\ arcmin$ all around the image are mostly due to some problems in the vignetting correction at the very edge of the field.

The emission from the cluster, which is completely contained in the central 20 $arcmin$ radius region, appears to be elongated in the direction from North–West to South–East, as already noticed in the optical data (see Paper I). A substantial number of discrete sources is also visible in the same area.

The most prominent features outside the ribs are two areas of enhanced emission, eastward of the cluster and partly covered by one of the support ribs of the PSPC. The inner one appears to consist of two discrete sources surrounded by enhanced extended emission, while the outer one, near the eastern edge of the field, consists only of extended emission. These two areas of extended emission correspond to the two poor clusters $SC\ 1327-312$ and $SC\ 1329-313$ studied by Breen et al. (1994), in their analysis of an EINSTEIN observation pointed on $SC\ 1329-313$. Additional extended features appear near the edge of the field in the second, third and at the intersection between fifth and sixth octants. (Octant 1 is the northern one, with the others following in clockwise direction). Although not clearly visible in this reproduction, the excess of counts in these areas is quite significant. The position of the second of these sources coincides with the center of $A3556$, while the other two sources correspond quite closely to the positions of minor enhancements in the optical counts of galaxies (see figure 1 in Paper I).

In order to estimate the background counts to be used in the following analysis, we have extracted the counts in the eight circular regions shown in Figure 1. These circles, with radii between 6 and 8 *arcmin*, are offset by roughly 40 *arcmin* from the image center. The size and position of these circles have been chosen in such a way to avoid as much as possible contamination from obvious discrete sources. The count rates corrected for vignetting and the corresponding errors for the eight circles are reported in Table 1, where the numbers given in the first column correspond to the various octants as indicated above. Inspection of the table clearly shows that circles #6 and #7 present a significant count excess, probably due to contamination from the emission of the two poor clusters $SC\ 1327-312$ and $SC\ 1329-313$ discussed above. These two regions have not been used in computing the average background. The average from the other six circles is 2.05 *counts/pixel* with a dispersion of 0.13, corresponding to $\sim 6\%$ of the average. The corresponding error on the mean is therefore of the order of 3%. Note that the dispersion of the values in the six circles, although small, is higher than what would be expected on the basis of purely Poissonian errors. Probably this is mainly due to large–scale disuniformities in the efficiency of the detector not properly corrected for, but it can also be in part due to the very complex structure of the X–ray emission of this particular region.

## 3. SPATIAL ANALYSIS

### 3.1 Determination of the spatial parameters of A3558

Figure 2 shows a contour plot of the whole field in which all the features described in the previous section are clearly seen. In this section we concentrate on the surface brightness distribution of *A3558*. This distribution has a roughly elliptical shape with a superimposed emission peak approximately corresponding to the position of the dominant galaxy (see below). For this reason, and given the very high statistics in the image, we have attempted a global fit to the surface brightness distribution through a function which is the sum of an elliptical King law (to fit the cluster), a Gaussian (to fit the central source) and a constant term (to estimate the background). The model function is therefore:

$$P(x,y) = I_{cl}\left[1 + \left(\frac{x'}{R_1}\right)^2 + \left(\frac{y'}{R_2}\right)^2\right]^{-3\beta+0.5} + I_g \exp\left[\frac{(x-x_g)^2 + (y-y_g)^2}{2\sigma^2}\right] + bck \quad (1)$$

where

$$x' = (x - x_{cl})\cos\theta + (y - y_{cl})\sin\theta \qquad (2)$$

$$y' = -(x - x_{cl})\sin\theta + (y - y_{cl})\cos\theta. \qquad (3)$$

Note that, in order to be consistent with the ROSAT image, we have adopted a right hand coordinate system with the origin in the upper left corner and the y-axis increasing downward. With this coordinate transformation, in which $\theta$ is a clockwise angle, the $x'$ axis is along the direction of the minor axis, and the $y'$ axis is along the direction of the major axis. The resulting position angle of the cluster, defined as usually as the direction of the major axis from North through East, is therefore given by P.A. = $\theta$ + 90 degrees.

The 12 variables to be simultaneously estimated are the two normalizations $I_{cl}$ and $I_g$, the positions $(x_{cl}, y_{cl})$ and $(x_g, y_g)$ of the centers of the cluster and of the galaxy respectively, the position angle of the cluster $(\theta)$, its minor and major core radii ($R_1$ and $R_2$), the exponent of the King law ($\beta$), the width of the gaussian ($\sigma$) and the background level ($bck$). In order to find the best fit parameters we minimized the quantity

$$\chi^2 = \sum_i \chi_i^2 = \sum_i \frac{(O_i - P_i)^2}{(P_i + (0.1 P_i)^2)} \qquad (4)$$

where $O_i$ and $P_i$ are the observed and predicted counts in the $i^{th}$ pixel, respectively. Apart from the Poissonian noise, we added a further 10% error on the predicted counts in order to take into account the pixel to pixel fluctuations due to instrumental effects (Hasinger et al. 1993).

Since a $\chi^2$ analysis can not be applied when the expected counts in each bin are small (i.e. $\leq$ 5–10), the original 15 × 15 $arcsec^2$ pixels were rebinned in groups of 3 × 3. In this way, the background counts by themselves, without any cluster contribution, are already of the order of $\sim$ 18 in each "binned pixel" (see Section 2). In order not to be affected by the ribs, we restricted our analysis to a circle of 17.5 $arcmin$ radius around the center.

While the very central region of the analyzed image is largely dominated by the emission of the cluster, the most external regions may be substantially contaminated by the presence of sources unrelated to the cluster. In order to take into account these sources, we have adopted an iterative fitting procedure. At each iteration in the minimization of eq.(4) we have identified the most "deviant" pixels, which have then been eliminated, together with the immediately adjacent pixels, in the next iteration. This procedure is rapidly converging;

after a few iterations we found the best fit parameters given in Table 2, with a reduced $\chi^2 = 1.22$ over a total of 1515 pixels. Notice that in this table the spatial parameters are expressed in units of original pixels (i.e. $15 \times 15$ $arcsec^2$). The total number of pixels eliminated from the final fit is 194, corresponding to 15 contaminating sources or "possible" sources. Each of these sources has at least one pixel with $\chi^2 > 15.0$ with respect to the model with the final best fit parameters. Among the 1515 pixels used in the final fit, the total number of pixels with $\chi^2 > 9$ is 6, very well consistent with the number expected on the basis of purely statistical fluctuations around the fit. With the exclusion of four of these "sources", which are located at the eastern edge of the analyzed region, where there is a clear indication of enhanced diffuse emission (see Figure 2), some parameters (position, counts, flux) for the other 11 sources found with this procedure are listed in Table 3 (see Section 3.3 below). The nominal best fit position of the central emission peak (see $x_g$ and $y_g$ in Table 2) is $\sim 20$ $arcsec$ from the optical center of the cD galaxy [$\alpha(2000) = 13^h27^m56^s.7$, $\delta(2000) = -31°29'45''.8$; Metcalfe et al. 1987], but well inside its optical extension ($R_{b_{26.75}} \sim 55$ $arcsec$).

In order to test the significance of the relative distance between the optical and X–ray peaks, we have repeated the global fit (cluster+galaxy) by fixing the position of the Gaussian at the optical position of the cD galaxy. On the basis of the resulting $\chi^2$ value, we find that the X–ray position coincident with the optical center of the cD galaxy is still acceptable at $\sim 2$ sigma confidence level. Moreover, from an analysis of a public HRI map, where *A3558* is offset by 10 *arcmin* with respect to the image center, it appears that the diffuse emission around the cD is centered at the optical position, but with a secondary emission peak consistent with our best fit position. On the basis of this analysis, from now on we will assume that the X–ray emission peak is associated with the cD.

An exact estimate of the errors for all the parameters listed in Table 2 would require a long and complex minimization procedure in the 12–dimensional space of the parameters. In order to obtain at least an approximate estimate of these errors we have followed two different procedures. First, we have checked the robustness of our best fit parameters by varying the area over which we have applied our $\chi^2$ analysis. For any circle with a radius greater than 12.5 *arcmin* all the best fit parameters given in Table 2 are essentially unchanged, with maximum variations of the order of 2%; restricting even further the radius of the analyzed region, the only parameters which are somewhat modified are the core radii, the $\beta$ parameter and the $\sigma$ of the Gaussian. For example, for a radius of 9 *arcmin* both the core radii and the $\beta$ parameter decrease by about 13%, while the $\sigma$ of the Gaussian increases of the same amount. All the other parameters do not vary by more than a few percent. In particular, the position angle varies by less than 4 degrees and it is interesting to note that it is in excellent agreement with the position angle derived from the distribution of galaxies ($\theta_{opt} = 130^o \pm 15^o$; Metcalfe et al. 1994). Secondly, we have computed the errors only for the parameters which will be used later on in deriving physical quantities (i.e. $I_{cl}$, $I_g$, $\beta$, $R_1$, $R_2$ and $\sigma$), keeping fixed the others at their best fit values. The $1\sigma$ errors for these parameters have been derived by constructing a grid of values and considering only those for which $\chi^2 - \chi^2_{min} < 7$, where 7 is the value for which the integral $\chi^2$ distribution with 6 degree of freedom is equal to 0.32. For all the parameters the resulting $1\sigma$ errors are of the order of $2 - 4\%$. Moreover, the fitted value for the background (2.014 $counts/pixel$) can be directly compared to the average background counts obtained from the 6 reference circles as described in the previous section. Although the two values have been obtained in two completely independent and different ways, they are perfectly consistent with each other, with a difference which is less than 2%.

Given this estimate on the errors, we can conclude that the central source, which has a gaussian dispersion of $37 \pm 1$ $arcsec$, corresponding to a FWHM of $87 \pm 3$ $arcsec$, is not consistent with being a point source. In fact, at the center of the field the FWHM is $\sim 30$ $arcsec$ at the average energy of 0.5 $keV$, while even at the lowest energy bin of 0.14 $keV$ the FWHM of the point spread function (54 $arcsec$) is significantly smaller than our best fit value. The difference between the positions of the central emission peak and the global center of the cluster emission ($\sim 1.3$ $arcmin$) appears to be highly significant. The $\chi^2$ value obtained with the constraint that the two centers coincide is not acceptable at more than $5\sigma$ confidence level. This displacement between the cluster isophotes and the central source is also clearly visible in Figure 2.

Figure 3 shows the radial distribution of the background subtracted surface brightness in circles centered around the peak of the X-ray emission. The dotted line shows the contribution of the King law fit, while the solid line shows the total model contribution (King law + central gaussian source). Except for the last point the sizes of the statistical errors are approximately of the same size as the symbols. The figure shows the goodness of the fit up to 17.5 $arcmin$: all the data points are within a few percent from the model. The largest discrepancy between model and data is in the third and fifth bin, at $\sim 1$ $arcmin$ and 2 $arcmin$ respectively from the center, in which the model differs of $\sim 8\%$ from the data. This suggests that probably the gaussian shape which we used to represent the surface brightness distribution of the central source has too much power in the tail.

### 3.2 Determination of the physical parameters

From the normalization of the King and Gaussian functions, it is possible to obtain the central surface brightness, using the appropriate conversion factor from counts to intrinsic flux (i.e. corrected for absorption). Assuming a Bremsstrahlung emission from a hot gas with a temperature of 3.25 $keV$, an abundance of 0.32 and a hydrogen column density of $3.84 \times 10^{20}$ $atoms$ $cm^{-2}$ (see Section 4.1), we find the following conversion: 1 PSPC $count$ $s^{-1}$ in $[0.14 - 2.13]$ $keV = 1.09 \times 10^{-11}$ $erg$ $s^{-1}$ $cm^{-2}$ in $[0.5 - 2.0]$ $keV$. In the rest of the paper, although we have used photons from a larger energy band, we will quote fluxes and luminosities in the more standard $[0.5 - 2.0]$ $keV$ band. As noted by Bower et al. (1994), all temperatures in the range $[2 - 10]$ $keV$ give conversion factors that differ by less than 3%: indeed the conversion factor for the cD galaxy ($kT = 1.89$ $keV$, abundance $= 0.61$) is $1.07 \times 10^{-11}$.

The two resulting central surface brightnesses in the $[0.5 - 2.0]$ $keV$ band are $I_{cl} = 5.2 \times 10^{-6}$ $erg$ $cm^{-2}$ $s^{-1} sr^{-1}$ and $I_g = 1.0 \times 10^{-5}$ $erg$ $cm^{-2}$ $s^{-1} sr^{-1}$. From $I_{cl}$ it is in principle possible to derive, with some assumptions on the intrinsic geometry of the system, the central emissivity $\epsilon_{cl}$. Since the elliptical distribution of the surface brightness of the cluster rules out a simple spherical symmetry, we have derived the central emissivity $\epsilon_{cl}$ in four extreme cases: prolate and oblate ellipsoids, with symmetry axis parallel and perpendicular to the line of sight. For all these cases the central emissivity can be written as:

$$\epsilon_{cl} = \frac{I_{cl}}{\sqrt{\pi} R_i} \frac{\Gamma(3\beta)}{\Gamma(3\beta - 0.5)} \quad (5)$$

where $\Gamma$ is the Euler complete Gamma function and $R_i$ is the major axis in the case of oblate ellipsoid with symmetry axis parallel to the line of sight and prolate ellipsoid with symmetry axis perpendicular to the line of sight. In the other two complementary cases, $R_i$ is the minor axis. For the spherical case, we assume a radius which is the geometrical mean between the major and the minor radii, i.e $R_i = \sqrt{R_1 \times R_2}$. We obtain $\epsilon_{cl} = (5.8^{+1.01}_{-0.85} \times 10^{-30})$ $h$ $erg$ $cm^{-3}$ $s^{-1} sr^{-1}$, where the central value corresponds to the spherical case and the upper and lower values correspond to the ellipsoid cases.

For the cD galaxy, the deprojection of a symmetric Gaussian function is

$$\epsilon_g = I_g / \sqrt{2\pi\sigma^2} \quad (6)$$

and we obtain $\epsilon_g = 49.5 \times 10^{-30}\ h\ erg\ cm^{-3}\ s^{-1} sr^{-1}$.

Following Henry et al. (1993), it is possible to estimate the central electron density from the Bremsstrahlung emission as

$$n_e = 1.60 \times 10^{12} \frac{\epsilon^{0.5}}{(kT)^{0.25}} \left[\gamma(0.7, E_2/kT) - \gamma(0.7, E_1/kT)\right]^{-0.5} \qquad (7)$$

where the gaunt factor was approximated by $g(E,T) = 0.9\,(E/kT)^{-0.3}$ and $\gamma(a,b) = \int_0^b e^{-t} t^{a-1} dt$ is the incomplete gamma function. For the temperature we adopted the value $kT = 3.25\ keV$, which is the average temperature of the cluster excluding the emission around the dominant galaxy (see Section 4.1). Note that the difference between the Henry et al. (1993) formula and eq.(7) is that we use the deprojected quantity $\epsilon$. With these parameters we obtain $n_{e,cl} = (4.3^{+0.4}_{-0.3}) \times 10^{-3}\ h^{0.5}\ cm^{-3}$ for the values of $\epsilon_{cl}$ derived from eq.(5), and $n_{e,g} = 13.6 \times 10^{-3}\ h^{0.5}\ cm^{-3}$ for the central Gaussian, for which the adopted temperature is $kT = 1.89\ keV$ (see Section 4.1).

As a first approximation, we estimated the observed flux and luminosity of the cluster directly from the total number of net counts detected within the central 20 $arcmin$. Strictly speaking, the values derived in this way are a slight overestimate (of the order of a few percent) of the real quantities, because the contributions from the dominant galaxy and other sources in the field (see Section 3.3 below) are not subtracted, but have the advantage of being directly comparable with previous determinations. The resulting flux and luminosity are $F_X = 4.4 \pm 0.4 \times 10^{-11}\ erg\ cm^{-2} s^{-1}$ and $L_X = 1.1 \pm 0.1 \times 10^{44}\ h^{-2}\ erg\ s^{-1}$, in the energy range $[0.5 - 2.0]\ keV$; the quoted 10% errors are essentially due to uncertainty in the absolute calibration of the PSPC (Briel et al. 1991).

Scaling the previous luminosity determinations for *A3558* to our energy range, we find that $L_{EXOSAT}$, $L_{GINGA}$ and $L_{Einstein}$ are 1.1, 2.2 and $0.8 \times 10^{44}\ h^{-2}\ erg\ s^{-1}$, respectively (Edge & Stewart 1991a; Day et al. 1991; Breen et al. 1994). The large discrepancy of $L_{GINGA}$ with respect to the other measurements is probably due the large field of view of GINGA ($1^o \times 2^o$ FWHM), which may result in significant contamination from nearby sources. All the other measurements are in excellent agreement with each other. Note that the Einstein luminosity was derived from the slew survey within a radius of $0.5\ h^{-1}\ Mpc$; in the same region we obtain $L_X = 0.8 \times 10^{44}\ h^{-2}\ erg\ s^{-1}$.

As an additional check on the results of the surface brightness spatial fit we have also computed the luminosity from our analytical radial fit by integrating the volume emissivities derived above. Performing the integral over a sphere of 0.8 $h^{-1}$ $Mpc$ (corresponding to 20 $arcmin$) we find $L_{fit} = 1.1 \times 10^{44}$ $h^{-2}$ $erg$ $s^{-1}$, in perfect agreement with the value obtained by summing all the counts. The fitted luminosity for the cD galaxy is $L_g = 5.1 \times 10^{42}$ $h^{-2}$ $erg$ $s^{-1}$, which corresponds to $\sim 5\%$ of the cluster luminosity. This value is typical for a giant cluster galaxy (Fabbiano 1989).

### 3.3 Other sources in the field

As already mentioned, detection of additional sources, not related to the cluster, in the central 20 $arcmin$ is complicated by the presence of the strong cluster emission. Following the procedure described in Section 3.1 we created a $\chi^2$ map with respect to the model describing the emission from the cluster and the central galaxy. Figure 4 shows this map for the innermost 20 $arcmin$ radius field, where a number of sources is clearly visible. The circles around these $\chi^2$ enhancements represent the areas from which the photons have been extracted in order to estimate the flux. The pixel size in this map is 45 $arcsec$. From this figure it is clear that the cluster and its dominant galaxy are well subtracted. Moreover, in addition to the sources indicated in the map, it is clearly evident the presence of a significant diffuse emission in the eastern part of the map, which is related to the poor cluster $SC$ $1327 - 312$ (see Figure 1 and Section 2). For the outer part of the map, where the cluster emission is not prominent, we chose to list the eight sources which have at least two contours in Figure 2; two of these sources (#18 and #22) correspond to the two poor clusters already discussed.

Table 3 reports some data for these additional sources, sorted by increasing right ascension. Column (1) is a sequential number, column (2) and (3) are the coordinates (2000), column (4) is the distance from the map center (in $arcmin$), column (5) and (6) are the net counts and the flux. For the conversion from counts to flux we have adopted the same conversion factor used for the cluster. This table contains 22 sources, 14 of which are within the circular support structure at $\sim 20$ $arcmin$.

The positions of these sources have been determined by fitting a gaussian in the $15 \times 15$ $arcsec^2$ pixel size image, using as a first guess the center obtained from the binned image. The pixel positions have been converted to equatorial coordinates using the TRANSFORM/COORDINATES program of EXSAS package. For source #4, partially covered by the circular rib at $\sim 20$ $arcmin$ from the center, we quote the position of the emission maximum. This procedure leads to coordinates which are within few arcseconds from those obtained with the standard maximum likelihood analysis of EXSAS.

In order to have a rough estimate of the flux of these sources, for all of them, except for the two identified with the poor clusters, we integrated the counts in the circle whose radius would contain 90% of the light for a point source in the lowest used energy channel (i.e. channel 14). This means that more than 90% of the photons over the entire band are included in such circle, with the exact percentage being a function of the energy distribution of the sources. For the two poor clusters we used a larger integration box. In both cases, however, the flux estimates are likely to be lower limits: $SC\ 1327-312$ is partly obscured by the rib, while $SC\ 1329-313$ is probably not completely contained inside the image. The agreement of our flux estimate for $SC\ 1329-313$ with the value reported by Breen et al. (1994) is very good (within 4%), while for $SC\ 1327-312$ our value is lower of $\sim 13\%$, after having subtracted the contribution of the adjacent source #20. This difference could arise from the unfortunate position of this group in both the ROSAT and EINSTEIN images, nearby the ribs in the former case and at the border of the field of view in the latter. For the sources within 20 $arcmin$ we obtained the net counts by subtracting from the observed counts those predicted at the source location by the model for the cluster emission described in eq.(1). For the sources outside 20 $arcmin$, instead, we subtracted a local background estimated in the vicinity of each source.

For what concerns the source identification, the list has been cross correlated with the Simbad database and the COSMOS/UKST galaxy catalogue, limited to $b_J = 19.5$, finding four possible identifications, two with stars and two with galaxies. The two sources identified with stars are source #4 at $\sim 17$ $arcsec$ from $LH2739$, an M star with $V = 13.6$, and source #20 at $\sim 4$ $arcsec$ from $HD\ 117310$, also known as $SAO204527$, a $K1\ III$ star with $V = 10.2$. The two sources identified with bright galaxies are source #6 at $\sim 7$ $arcsec$ from a galaxy with $b_J = 15.1$ and source #9 at $\sim 4$ $arcsec$ from a galaxy with $b_J = 15.8$. The radial velocity of these galaxies is $13,186$ and $12,801$ $km/s$ respectively (Teague et al. 1990), and therefore they are inside the velocity range covered by the galaxies in the core of the Shapley Concentration (see figure 3.b in Paper I). For all these four sources the ratio of X−ray to optical flux is well inside the range of typical values of X−ray selected M and K stars and galaxies, respectively (Ciliegi et al. 1994).

These identifications show that our position determination is good and no boresight correction is needed. Indeed, apart from $LH2739$, whose relatively large distance from the optical counterpart is probably due to the difficulty in the determination of its X−ray position because of the proximity of the rib, the positional differences of the other three X–ray sources with optical counterparts are well consistent with expected errors of $\sim 6$ $arcsec$ reported by Küster & Hasinger (1993).

Sources #18 and #22 are located at the positions of the poor clusters $SC$ $1327-312$ and $SC$ $1329-313$ respectively: the distances of our sources from the centers given by Breen et al. (1994) are $\sim 26$ $arcsec$ and $\sim 2$ $arcmin$ respectively. Since, as already mentioned, it is difficult to derive precise positions from our data for these two sources, we consider satisfactory this positional agreement.

Very close to the center of source #18 ($\sim 7$ $arcsec$) there is also a bright galaxy, with $b_J = 15.6$, which is likely to be the dominant galaxy of this poor cluster and may contribute significantly to the observed emission.

As expected from their location in cluster and/or group environment and from their relatively strong X–ray emission, the colors (Metcalfe et al. 1994) of the three galaxies we found as possible optical identifications suggest that all of them are early–type galaxies. This is confirmed also by a visual inspection of the Palomar prints.

## 4. SPECTRAL ANALYSIS

### 4.1 Temperature estimates

In order to perform the spectral analysis we corrected all the spectra by dead–time, vignetting and effective area using the standard correction procedures offered by EXSAS, as mentioned in Section 2. Due to the loss of gain of the PSPC (ROSAT News N.27), the choice of the correction matrix is not unique, but there are two *Detector Response Matrixes*, named $drmpspc_{06}$ and $drmpspc_{36}$, nominally corresponding to observation periods before and after October 1991, respectively. Because this date is only indicative (the loss of gain changes continuously with time), we will use in our analysis the "old" matrix $drmpspc_{06}$, but we will present the results of the spectral analysis also using the "new" matrix $drmpspc_{36}$.

In order to study the radial dependence of the spectral properties, the cluster was divided into six circular annular regions around the fitted position of the center (see Table 4). In each of these regions the cluster counts, after background subtraction, are at least of the order of 10,000. For the background we adopted the mean value derived from the six reference circles discussed in Section 2. Before deriving the spectral parameters of the cluster, it is necessary to subtract the contribution of the dominant galaxy. For this reason, in the first two annuli we eliminated a sector whose axes are tangent to a circle centered on the galaxy and with radius of $1.5\sigma$ (where $\sigma$ is the dispersion of the Gaussian): this corresponds to an opening angle of $50°$. We eliminated also the contribution from the discrete sources described in Section 3.3, excluding the counts inside 1 FWHM of the their $PSF$ at channel 7 (the typical radius is $\sim 1.3$ $arcmin$). However, even without subtraction of the counts from discrete sources, the results would change typically by 10% at most.

The spectral parameters of the dominant galaxy were estimated by using the counts within a circle centered on its fitted position and with a radius of 0.83 $arcmin$ (corresponding to $1.3\sigma$). In order to account for the contribution of the cluster emission, we subtracted from these data the counts derived in a similar circle, placed in a symmetric position with respect to the cluster center.

The spectra were analysed with the XSPEC package (provided by the NASA/HEASARC). The spectral channels have been rebinned in such a way that in each bin the signal to noise ratio is at least 10. We adopted a model consisting of emission from a hot, diffuse plasma (Raymond & Smith 1977), absorbed by a galactic neutral hydrogen column and we first fitted the temperature, the spectrum normalization, the metal abundance and the hydrogen column density using counts in the energy range $[0.14 - 2.13]$ $keV$. The weighted mean of the best fit values of the hydrogen column density in the six annuli is $3.84 \pm 0.20 \times 10^{20}$ $atoms$ $cm^{-2}$, in good agreement with the value reported in the EXSAS database ($3.63 \times 10^{20}$ $atoms$ $cm^{-2}$). From the Bell Laboratories HI Survey (Stark et al. 1992, privately distributed tape) we obtained an $N_H$ value of $4.35 \times 10^{20}$ $atoms$ $cm^{-2}$. Given this discrepancy between these two non–X–ray determinations, we fixed $N_H$ at the best fit value derived from our data and we repeated again the fits for each annulus.

In Table 4a and 4b we report the best fit parameters of the various regions obtained using the $drmpspc_{06}$ and $drmpspc_{36}$ calibration matrixes, respectively. Column (1) represents the chosen region, column (2) gives the background subtracted counts of the spectra, column (3) is the estimated temperature (in $keV$) and its one sigma error, column (4) is the metal abundance, normalized to solar, with its error, and column (5) represents the reduced $\chi^2$ of the fit.

The temperature for the first 5 bins (corresponding to a region of 13 $arcmin$ radius from the cluster center) is consistent with an isothermal distribution and also the abundances do not show any radial trend. The best fit value of the temperature in the last bin (2.23 $keV$;

triangle with dashed error bars in Figure 5a) is slightly lower than the temperature in the inner regions. It is however difficult to assess the reality of this determination, because in this bin the contribution of the background counts ($\sim 18,000$) with respect to those from the cluster ($\sim 10,000$) is relatively high, and it is in principle possible that the results of the fit are somewhat affected by residual systematic uncertainties in the background correction. To test this possibility we fitted the last bin also by subtracting separately the spectra of the six background regions, but the results appeared to be reasonably stable. Another possible reason for the lower temperature obtained in this bin is the contamination of the cluster emission by the diffuse emission (see Figure 2 and Section 3) probably due to the poor cluster $SC\ 1327-312$. We have therefore eliminated from the fit a sector of 45 degrees around this feature, remaining with $\sim 7,000$ background subtracted counts. Because of the reduced number of counts, in order to have a better determination of the temperature, we fixed the abundance at the average value of the other five annuli (i.e. 0.32), obtaining a temperature of $3.06^{+1.64}_{-0.91}\ keV$ (circle with solid error bars in Figure 5a). This value is now consistent with the others and therefore we conclude that the derived temperature profile is statistically consistent with an isothermal distribution of the gas in the cluster at least up to a distance of 18.4 $arcmin$, corresponding to 0.77 $h^{-1}\ Mpc$.

However, the fact that each of the first three radial bins has an higher temperature than the last three suggests the possible existence of a slight decreasing radial trend in the temperature profile. In order to test this hypothesis, we divided the cluster in two parts, with radius smaller and greater than 5 $arcmin$ (corresponding to the first three and the last three bins). The inner part of the cluster has $kT = 3.32^{+0.38}_{-0.31}\ keV$ and abundance $0.34^{+0.12}_{-0.13}$, while for the outer region $kT = 2.88^{+0.31}_{-0.30}\ keV$ and abundance $0.25^{+0.12}_{-0.12}$. Note that, even if these values are compatible with each other within the errors, their confidence ellipses overlap only at $2\sigma$ level. Therefore, although not required by the data, a radial decrease of the gas temperature could be present. In any case, the change in temperature, if present, is small enough that we can safely assume, in the following analysis, the reference values of $kT = 3.25\ keV$ and abundance 0.32, obtained as an average of the single values in the first five bins (see dotted lines in Figures 5a and 5b). The average temperature is well consistent with the value $kT = 3.7^{+1.0}_{-2.0}\ keV$ derived by Edge & Stewart (1991a) from EXOSAT data.

Using the same ROSAT data, Breen & Raychaudhury (1994) reported a mean temperature for $A3558$ of 4.5 $keV$, derived by fixing the abundance at the cosmic value. With such an abundance, we obtain a temperature of $3.84 \pm 0.29$ $keV$, significantly lower than their value. We do not know the reason for this difference. The discrepancy could arise from a different choice of the regions where the background is evaluated; moreover, Breen & Raychaudhury do not specify which value they adopted for $N_H$, and therefore a direct comparison with our results is not possible. However, in order to obtain $kT = 4.5$ $keV$ from our data we would need a column density of $N_H \sim 3 \times 10^{20}$ $atoms\ cm^{-2}$, significantly smaller than all the existing determinations. With such $N_H$ value the reduced $\chi^2$ (2.37) obtained from our fit is significantly worse than that obtained with our adopted parameters (1.27).

In Figures 5a and 5b we show also (with squares and dashed lines) the best fit values obtained for the dominant galaxy, plotted at the conventional distance of 0 $arcmin$. The nominal best fit for the temperature (1.89 $keV$) is lower than the average cluster value, while the abundance ($\sim 0.6$) is higher than the mean value of 0.32.

The data shown in Figures 5a and 5b refer to the values reported in Table 4a. The radial behaviour of temperature and abundance would be essentially the same if we used the values obtained with the second response matrix (see Table 4b), although with systematic positive offsets of about 0.35 $keV$ in temperature and 0.17 in abundance. In order to check if these results depend on the specific fitting routine used, the spectra were fitted also with the RSMS program implemented in the EXSAS package. The obtained temperatures are very similar to those obtained with the XSPEC package (the difference is on average of $\sim 3\%$), while the resulting abundances are systematically lower of $\sim 30\%$.

A detailed spectral analysis has been applied only to two of the other sources detected in our observation, because of the relatively small number of counts. One is $SC\ 1327-312$ (source #18 in Table 3), for which we extracted all the counts in a circle with a radius of 2.25 $arcmin$, centered on the emission peak. The total number of background subtracted counts used to fit the spectrum is 3512. For this source we fixed the redshift and the hydrogen column density at the values used for $A3558$, because it is reasonable that this group is part of the core of the Shapley Concentration. The resulting temperature is not well determined (see Tables 4a and 4b), but its best fit value (3.85 $keV$) is in good agreement with the value of $\sim 3.75$ $keV$, measured by Breen et al. (1994) on EINSTEIN data (see their figure 3a).

The other source is $HD\ 117310$ (source #20 in Table 3): we fitted its spectrum with a Raymond–Smith model, fixing $N_H = 3.84 \times 10^{20}$ $atoms\ cm^{-2}$. The results are reported in Tables 4a and 4b: the derived temperature $kT = 1.10^{+0.11}_{-0.11}$ $keV$ is consistent with the values generally reported for hot stellar coronae (see f.i. Schmitt et al. 1990)

**4.2 The $\beta$ problem**

Under the simple assumption that both the gas and the galaxies in the clusters are isothermal and in hydrostatic equilibrium, within a common potential, one derives that $\rho_{gas} \propto \rho_{gal}^{\beta}$ (Cavaliere & Fusco–Femiano 1976), where the parameter $\beta$ can be estimated directly from spectroscopic observations, by measuring the temperature of the hot gas and the velocity dispersion of the galaxies:

$$\beta_{spec} = \frac{\mu m_p \sigma^2}{kT} \qquad (8)$$

where $\mu = 0.57$ is the mean molecular weight corresponding to a metal abundance of 0.3, $m_p$ is the proton mass, $\sigma$ is the velocity dispersion of galaxies and $kT$ is the temperature of the cluster gas. In the isothermal–hydrostatic model, $\beta_{spec}$ should be equal to $\beta_{ima}$ as derived from the fit of the X–ray surface brightness distribution (see Section 3.1).

For the velocity dispersion we adopt the value 1016 $km/s$, derived in Paper I from galaxies within 6 $arcmin$ from the cluster center. The chosen temperature is 3.42 $keV$, which is the mean value derived from the first three radial bins, corresponding to a radius of 5 $arcmin$. With these values $\beta_{spec} = 1.79$, to be compared with $\beta_{ima} = 0.611$, thus indicating the presence of a quite extreme "$\beta$ problem". While the value for $\beta_{ima}$ is quite close to the average value measured for other X–ray clusters (Jones & Forman 1984), the value of $\beta_{spec}$ is among the highest measured (see figure 18 in Edge & Stewart 1991b and figure 3 in Lubin & Bahcall 1993). In order to obtain $\beta_{spec} \sim \beta_{ima}$, we should have either a temperature of $\sim 11$ $keV$ (maintaining the observed velocity dispersion) or a velocity dispersion of $\sim 565$ $km/s$ (maintaining the observed temperature).

The same conclusion is reached by using the relation between the galaxy velocity dispersion and the cluster temperature [$\sigma = 332\ (kT)^{0.6}\ km/s$] obtained by Lubin & Bahcall (1993) for a sample of 41 clusters. With our measured temperature, the predicted value for the velocity dispersion is $\sigma = 694\ km/s$, indicating that for $A3558$ the ratio $\sigma/T$ is higher than the mean, at a level of 2.4 sigma.

Bahcall & Lubin (1994) claimed that the $\beta$ problem results mainly from assuming a radial distribution $\propto r^{-3}$ for the galaxy distribution at large radii, while the actual observed profiles appear to have a shallower slope. Fitting the galaxy profiles for a sample of clusters from the literature, they find an average radial distribution $\propto r^{-2.4}$ and give (see their eq.9) the correction factor for $\beta_{ima}$, in order to obtain a correct $\beta_{ima}^c$ to be compared to $\beta_{spec}$. With this correction, the $\beta$ problem almost disappears from their sample. We tried the same solution with our data, using the galaxy profiles found by Metcalfe et al. (1994) for *A3558*, but the highest value we obtain is $\beta_{ima}^c = 1.43 \cdot \beta_{ima} = 0.87$, and therefore the $\beta$ discrepancy remains.

The existence of such an extreme $\beta$ problem and the high value for $\sigma/T$ in *A3558* indicate that its measured velocity dispersion is too high and/or its measured temperature is too low. The first effect could be due to an overestimate of the velocity dispersion induced by the presence of substructures, as discussed by Edge & Stewart (1991b), or could be a consequence of a merging process, as shown by the simulations of Navarro et al. (1995). The second effect, which appears to be much less probable, could be induced by a cooling of the whole cluster, which however is not expected on the basis of the derived gas density, or by significant instrumental calibration errors, which may result in a wrong temperature determination. However, the good agreement between our temperature and the value derived from EXOSAT data seems to rule out the latter hypothesis.

## 5. MASS ESTIMATES

### 5.1 Mass estimates

Assuming that the gas is non-rotating, in hydrostatic equilibrium and with a spherical symmetry distribution, the total mass (dark + baryonic matter) is given by (see for example Sarazin 1988):

$$M_{tot}(<r) = -\frac{kT(r)}{G\mu m_p}\left[\frac{d\ln \rho_{gas}}{d\ln r} + \frac{d\ln T(r)}{d\ln r}\right] r \quad (9)$$

Since the temperature profile of *A3558* is consistent with an isothermal distribution at $kT = 3.25\ keV$ (see Section 4.1), we can neglect the temperature derivative in eq.(9). The radial density distribution of the gas has been derived in Section 3.1 and the total gas mass $M_{gas}(<r)$ can be obtained by directly integrating such a distribution.

In Table 5 we report the derived masses within spheres of different radii around the cluster center: column (1) gives the radius in $h^{-1}\ Mpc$, columns (2) and (3) are $M_{gas}$ and $M_{tot}$ (in $10^{13}\ M_\odot$) respectively, and column (4) gives the ratio $M_{gas}/M_{tot}$. Note that the values beyond a radius of about $0.75\ h^{-1}\ Mpc$ have been obtained by extrapolating to larger radii the radial distribution of the gas density derived from the inner regions. The values reported in this table are derived in the spherical symmetry hypothesis: however, the values obtained by taking into account the ellipsoidal shape of the cluster would not differ by more than a few percents.

The ratio $M_{gas}/M_{tot}$ is in the range $15 - 25\%$ $h^{-3/2}$ (see Table 5): this value is much higher than those reported by David et al. (1995) for a sample of 7 clusters (see their figure 4), whose mean value is $\sim 10\%$ $h^{-3/2}$. This result confirms the existence of the problem already outlined by White et al. (1993b) in their detailed discussion of the masses in the Coma cluster: baryonic matter constitutes a larger fraction of the total mass of rich galaxy clusters than is predicted by the standard CDM models of the Universe (with $\Omega = 1$). The most likely explanations suggested by White et al. (1993b) for this observational result are that either the Universe has low density or the standard interpretation of the element abundances is incorrect.

Metcalfe et al. (1987) derived the optical blue luminosity for the innermost 5 $arcmin$ (corresponding to 0.2 $h^{-1}$ $Mpc$) of $A3558$, finding a value of $3.7 \times 10^{11}$ $h^{-2}$ $L_\odot$ and a corresponding mass–to–light ratio of $M/L = 392 \pm 120$ $h$ $M_\odot/L_\odot$. The total mass derived in the same region from our data is $2.4 \times 10^{13}$ $h^{-1}$ $M_\odot$, corresponding to $M/L = 65$ $h$ $M_\odot/L_\odot$, significantly lower than the Metcalfe et al. (1987) value. Metcalfe et al. (1994) present also the integrated blue luminosity as a function of the distance from the cluster center up to 1.75 $h^{-1}$ $Mpc$ (see their figure 10): with these data and our masses we find that the $M/L$ ratio rapidly increases up to $\sim 0.4$ $h^{-1}$ $Mpc$ and then converges to a value of $\sim 130$ $h$ $M_\odot/L_\odot$. Probably, the main reason for the difference in the $M/L$ values resides in the fact that Metcalfe et al. (1987, 1994) derived the mass on the basis of a virial estimator. Indeed, virial masses are often higher than the dynamical masses derived from X–ray data (see f.i. Mushotzky 1993 and David et al. 1995). Biviano et al. (1993) derived for $A3558$ a mass $M_{vir} = 6 \times 10^{14}$ $h^{-1}$ $M_\odot$, within 0.75 $h^{-1}$ $Mpc$, while Metcalfe et al. (1994) derived $M_{vir} = 1.2 \times 10^{15}$ $h^{-1}$ $M_\odot$ within 1.5 $h^{-1}$ $Mpc$. By comparing these numbers with our estimates reported in Table 5, it is seen that in both cases the estimates obtained from the virial theorem are a factor four higher than those derived from the X–ray data. This discrepancy could be related to the $\beta$ problem discussed in Section 4.2: if the measured velocity dispersion were significantly enhanced by the presence of substructures and/or recent merging, also the virial mass would be overestimated. Alternatively, if the kinetic pressure of bulk motions in the gas is not negligible with respect to the total pressure support, the total mass derived from the X–ray data could be underestimated up to $\sim 30\%$ (Evrard 1990, Navarro et al. 1995).

From the radial luminosity distribution (Metcalfe et al. 1994), it is possible to obtain the integral stellar mass through the mean $(M/L)_{star}$ ratio: we adopt the value $(M/L)_{star} = 6.4\,h$ reported by White et al. (1993b) and we compute $M_{star}$ as a function of the distance from the cluster center. Having also the radial distribution of the mass of the hot gas $M_{gas}$ and of the total dynamical mass $M_{tot}$, it is possible to study, at least qualitatively, the different behaviours of the various matter components. In the four panels of Figure 6 we show the ratios of the masses of the various matter components as a function of the radius. Since the masses of the different components do not scale in the same way with the Hubble constant, we show the results for both $H_o = 100\,km\,s^{-1}\,Mpc^{-1}$ (filled circles and solid lines) and $H_o = 50\,km\,s^{-1}\,Mpc^{-1}$ (open circles and dotted lines). The circles in the panels correspond to the radii at which Metcalfe et al. (1994) give the integrated cluster luminosity; we have not used their first data point, at a radius of $\sim 0.125\,h^{-1}\,Mpc$, because it is strongly contaminated by the cD galaxy.

The main conclusions we can reach from inspection of the four panels of Figure 6 are:

**a)** The hot gas component is more extended than both the dark and the stellar component [panels a) and b)]. This is in agreement with similar results obtained recently by Durret et al. (1994) and David et al. (1995).

**b)** The dark and the stellar components have approximately the same radial distribution at radii greater than $\sim 0.5\,h^{-1}\,Mpc$, for $h = 1$, but the stellar component is more extended than the dark matter for $h = 0.5$ [panel c)].

**c)** As already noticed by Durret et al. (1994), the stellar mass appears to be a good tracer of the total mass distribution, with an approximately constant ratio of $M_{tot}/M_{star} \sim 20$ for radii greater than $\sim 0.5\,h^{-1}\,Mpc$ [panel d)]. Note that in this panel the curves for the two $h$ values coincide.

**d)** For both $h = 1$ and $h = 0.5$ the mass in stars is the least important component. For $h = 1$ most of the mass is in dark matter ($M_{dark} : M_{gas} : M_{star} \sim 15 : 5 : 1$ within one Abell radius), while for $h = 0.5$ most of the mass is in hot gas ($M_{gas} : M_{dark} : M_{star} \sim 14 : 6 : 1$). As already mentioned, the values of $M_{gas}$ and $M_{tot}$ at one Abell radius have been obtained extrapolating to larger radii the radial distribution of the gas density derived from the inner regions. However, even if we limit ourselves to the inner region where we have measured data ($r \sim 18\,arcmin$), for $h = 0.5$ the mass in gas would be more than 50% of the total mass.

**5.2 The central cooling flow**

The presence, in the central part of a cluster, of a region of higher density and lower temperature may reveal the existence of a cooling flow (see f.i. Fabian et al. 1991 and references therein). Recently, Briel et al. (1991) and Fabian & Daines (1991) reported the existence of a cooling flow in *A2256*, but suggested that in this case this phenomenon might be due to a group which is merging with the cluster, because of the distance between the density spike and the geometrical center of the cluster ($\sim 175\ h^{-1}\ kpc$). In fact, if a merging group presents a cooling flow, it has a central density high enough to resist to the disruption by ram pressure, while its external gas is stripped away; moreover, the cooling gas will not necessarily be reheated (Fabian & Daines 1991). The existence of a group of galaxies merging toward the center of the cluster was suggested also by White et al. (1993a) for the Coma cluster, even if in this case there is no evidence for a cooling flow.

We derived in previous sections that near the center of *A3558* there is the presence of both a spike of density, associated with the cD galaxy, and a decrease of the temperature: therefore a cooling flow can be present there. In order to verify this hypothesis, we derive the cooling time, which can be estimated as (Henry & Henriksen 1986)

$$t_{cool} = 6.0 \times 10^3\ yrs\ \frac{T^{1/2}}{n_e} \qquad (10)$$

where $t_{cool}$ is the cooling time (in years), $T$ is the temperature (in $K$) and $n_e$ is the electron number density (in $cm^{-3}$). If we consider only the cluster, without the contribution of the Gaussian density spike, we find for the central region $t_{cool} \sim 1 \times 10^{10}\ h^{-0.5}\ yrs$: therefore the cooling time for the cluster is comparable to the age of the Universe. On the other hand, using the density estimated for the central spike, we obtain $t_{cool} \sim 2 \times 10^9\ h^{-0.5}\ yrs$, which therefore would meet the time constraint.

However, with the present data it is not possible to assess if this flow is associated to the cluster or to an infalling group, because the center of the X–ray Gaussian density spike is not exactly at the center of *A3558*, but it is close to it ($\sim 50\ h^{-1}\ kpc$).

The luminosity $L_{cool}$ produced by the cooling process is usually derived by integrating the emissivity in the region of the cooling flow, which can be defined as the region where $t_{cool}$ is less then the age of the Universe (Fabian et al. 1991). Under the hypothesis that $L_{cool}$ is due both to radiation and pressure work as the gas enters in the cooling region, it is possible to have a rough estimate of the mass deposition rate $M_{cool}$ (Fabian et al. 1991)

$$M_{cool} = \frac{2}{5}\frac{\mu m_p}{kT}L_{cool} \qquad (11)$$

where $L_{cool}$ is the bolometric luminosity. Although eq.(11) ignores the gravitational work on gas, Stewart et al. (1984) pointed out that it is a reasonable approximation and that this formula tends to underestimate $M_{cool}$ for small cooling flows.

Assuming $n_e = n_{e,g}$ and $kT = 1.89\ keV$, we find that $t_{cool} \leq 1 \times 10^{10}\ h^{-1}\ yrs$ in a region with a radius of $\sim 50\ h^{-1}\ kpc$, deriving $L_{cool} = 4.9 \times 10^{42}\ h^{-2}\ erg\ s^{-1}$, in the range $[0.5 - 2.0]\ keV$; converting this value to bolometric luminosity, we obtain $M_{cool} \sim 25\ h^{-2}\ M_\odot\ yrs^{-1}$, in excellent agreement with the typical value derived by Edge et al. (1992, see their figures 1 and 3) for clusters with X-ray luminosity similar to that of $A3558$. Note that the value for $L_{cool}$ is highly uncertain because it is strongly dependent on the radial distribution of the gas in the spike, which we have assumed to be Gaussian. Given the uncertainties, the value of $L_{cool}$ is consistent with the total X-ray emission of the cD galaxy and represents $\sim 4\%$ of the total luminosity of the cluster.

## 6. SUMMARY AND DISCUSSION

The core of the Shapley Concentration is a remarkable structure formed by three close interacting ACO clusters, the most interesting of which is the richness 4 class cluster $A3558$; in this paper we presented the analysis of a deep (30,213 seconds of exposure time) PSPC ROSAT observation, centered on this object. The image of the cluster appears to be elongated in the direction from North-West to South-East, as already noticed in the optical data. Moreover, it is clear the presence of two additional extended emission regions, corresponding to the poor clusters $SC\ 1327 - 312$ and $SC\ 1329 - 313$: these two groups appear to be physically connected to the cluster through a bridge of hot gas, revealed as a diffuse emission.

We fitted the surface brightness profile with an elliptical King law for the cluster and a Gaussian profile for the cD galaxy, and we found that the position of the X-ray emission associated with the dominant galaxy is $\sim 50\ h^{-1}\ kpc$ away from the cluster center. From the spatial parameters we derived also the physical parameters of the cluster, such as the central electron density and the central emissivity. Finally, we individuated a list of other discrete sources present in the map.

From the spectral analysis, we found that the temperature profile of $A3558$ is consistent with an isothermal distribution at $kT \sim 3.25\ keV$, while the emission around the cD galaxy is cooler, at $kT = 1.89\ keV$. Our temperature estimate is consistent with the value $kT_{EXOSAT} = 3.7^{+1.0}_{-2.0}\ keV$ found by Edge & Stewart (1991a), while is significantly lower than $kT_{GINGA} = 6.2 \pm 0.3\ keV$ (Day et al. 1991): however the GINGA estimate is probably contaminated by other sources (see Section 2). Our derived metal abundance for the cluster is 0.32, well consistent with the typical values obtained for clusters with GINGA (Hatsukade 1989) and ASCA (Bautz et al. 1994) data.

The total luminosity inside $0.8\ h^{-1}\ Mpc$ for $A3558$ is $1.1 \times 10^{44}\ h^{-2}\ erg\ s^{-1}$ (in the range $[0.5 - 2.0]\ keV$), with a contribution from the cD galaxy of $\sim 5\%$. The derived luminosity is consistent within few percent with $L_{EXOSAT}$ and $L_{Einstein}$, while is significantly lower than $L_{GINGA}$ (see Section 3.2).

The total dynamical mass inside $1.5\ h^{-1}\ Mpc$ is $3.1 \times 10^{14}\ h^{-1}\ M_\odot$, with a $\sim 20\%\ h^{-3/2}$ component due to the hot gas. Combining these masses with the luminosity data from Metcalfe et al. (1994), we derived that the $M/L$ ratio is $\sim 130\ h\ M_\odot/L_\odot$, for radii larger than $\sim 0.4\ h^{-1}\ Mpc$. Studying the profiles of the three mass components (hot gas, dark matter and stellar mass), we concluded that the gas has the most extended distribution. For $h = 1$ most of the dynamical mass is in dark matter, while for $h = 0.5$ most of the mass is in hot gas. In both cases, the stellar mass appears to be a good tracer of the total mass distribution.

Finally, under the hypothesis that the central density spike represents a cooling flow, we derived a mass deposition rate of $\sim 25\ h^{-2}\ M_\odot\ yrs^{-1}$.

For what concerns the comparison with other X-ray emitting clusters, it results that the single parameters derived for $A3558$ are quite typical. Its temperature is consistent with, although on the low side of, the temperature distribution of the 46 clusters in the Edge & Stewart (1991a) sample (in which $A3558$ is not included). Its luminosity is slightly lower than the median X-ray luminosity of the $R_2$ clusters (i.e. clusters with richness $R = 2$), as derived by Burg et al. (1994) in their study of the X-ray luminosity function of Abell clusters. Because of the existence of a correlation between richness and X-ray luminosity, the fact that $L_X(A3558) < L_X(median)$ of $R_2$ clusters suggests the possibility that the optical richness class of $A3558$ may be overestimated, because of the contamination from nearby groups and clusters. Indeed Breen et al. (1994), comparing $A3558$ with the other six Abell/ACO clusters of richness class 4–5, found that only $A1146$ has a lower X-ray luminosity. This hypothesis is confirmed also by the photometric study of Metcalfe et al. (1994), who estimated for $A3558$ a corrected richness class 2.

The conclusion that the richness class of $A3558$ may be overestimated is supported also by the relatively small value of its total dynamical mass. In literature there are only a few good X-ray mass estimates: in particular for the four richness 2 clusters *A426* (Perseus), *A1656* (Coma), *A1795* and *A2256* Mushotzky (1993) reports a mass of 1.7, 2.3, 2.0 and $2.5 \times 10^{14}$ $h^{-1}$ $M_\odot$, respectively, within 0.5 $h^{-1}$ *Mpc*. All these values are about a factor two higher than the mass we derived for *A3558*.

Although its parameters are quite typical, *A3558* presents some peculiarities in the correlations between these quantities. First, its luminosity is too high with respect to its temperature: in fact, using the relation between temperature and total bolometric luminosity, fitted by Henry & Arnaud (1991) on a sample of 25 clusters, we find that the temperature of *A3558* corresponds to a bolometric luminosity ($6.2 \times 10^{43}$ $erg\ s^{-1}$) significantly lower than the measured value of $3.1 \times 10^{44}$ $erg\ s^{-1}$. This fact is related to the result that the ratio between the hot gas and the total dynamical mass is much higher ($\sim 20-25\%\ h^{-3/2}$) than the typical value of $\sim 10\%\ h^{-3/2}$ (White et al. 1993b, David et al. 1995). Both these results could be explained with an excess of baryonic matter in *A3558*, probably due to a contamination of this cluster by material accreted from nearby groups: such material would increase both the gas mass and the X-ray luminosity. An indication of possible merging processes in this cluster is also given by the presence of a quite extreme "$\beta$ discrepancy".

The fact that the hot gas is the most extended mass component is not a new result: evidences that the dynamical matter is more centrally concentrated than the hot gas are present also in other clusters (see Mushotzky 1993, Durret et al. 1994, David et al. 1995) and are confirmed by the results from the gravitational lens analysis, which suggest that the core radii of the total mass of clusters are a factor $2 - 2.5$ smaller than those derived from the optical and X-ray analysis (see Bartelmann & Narayan 1994 for a review). Navarro et al. (1995) explained the different profiles of dark matter and hot gas as a consequence of a merging process: analysing the behaviour of the two components during simulations of mergings, they found that the internal energy is transferred from the dark matter to the baryonic one, resulting in an expansion of the latter. As noted by Durret et al. (1994), a similar explanation holds also for the different behaviour between stellar and gas mass, given the fact that the stars are essentially collisionless (as the dark matter), while the gas is collisional. The hypothesis of a recent merger in *A3558* can be reinforced also by the fact that Navarro et al. (1995) found in their simulations that larger values of $\beta_{spec}$ occur in clusters which have undergone recent mergers.

Finally we note that *A3558*, although it is likely to be in interaction with other groups and clusters, show some evidence for the presence of a central cooling flow, with an estimated mass deposition rate in agreement with the typical values. This fact could indicate that either the interaction phenomenon was not able to disrupt the cooling flow or that the potential well of *A3558* managed to avoid the defocusing (MacGlynn & Fabian 1984). On the other hand, it is also possible that the cooling flow is not associated to the cluster but to an infalling group, which is merging with *A3558* (Fabian & Daines 1991).

A detailed discussion of the substructures in *A3558* and of the galaxy density profile, together with many new redshift measurements in the core of the Shapley Concentration, will be presented in a following paper (Bardelli et al., in preparation).

**Acknowledgements**
We thank Richard Burg for his help in preparing the ROSAT proposal. We warmly thank Andrea Comastri for his helpful assistance in the use of the reduction packages. SB acknowledges a three–month EEC fellowship at DAEC, Meudon Observatory, France. We also thank the referee for useful comments.

This work has been partially supported by the EEC grant ERB–CHRX–CT92–003. References retrieved from the Simbad database maintained by CDS, Strasbourg, France, were useful in preparing this paper.

# FIGURE CAPTIONS

**Fig.1**

A grey scale representation of the PSPC image of the cluster *A3558*. The whole PSPC field of view is shown, covering an area of the sky approximately 2 × 2 square degrees. The pixel size is $15 \times 15\ arcsec^2$; North is up and East is left. The cluster is the large, extended source in the center of the field; the other discrete sources are discussed in the text. The 8 numbered circles correspond to the region where the background counts were estimated (see Table 1).

**Fig.2**

A contour map of the emission in the whole PSPC field. The image has been obtained binning the data in $15 \times 15\ arcsec^2$ pixels, with a smoothing of $4 \times 4$ pixels. The origin $(0,0)$ of the coordinates corresponds to the nominal pointing of the PSPC image and the $X$ and $Y$ scales are in *arcsec*. The emission peak superimposed to the elliptical shape of the cluster corresponds to the dominant galaxy, while the numbered peaks are the discrete sources present in the outer part of the field (see Section 3.3).

**Fig.3**

Radial distribution of the background subtracted surface brightness in circles centered around the peak of the X–ray emission. The dotted line shows the contribution of the King law fit, while the solid line shows the total model contribution (King law + central Gaussian source). Except for the last point the sizes of the statistical errors are approximately of the same size as the symbols.

**Fig.4**

Map of the innermost 20 *arcmin* radius field, in which for each pixel it is reported its value of $\chi_i^2$; the pixel sizes are $45 \times 45\ arcsec^2$. It is clear that the cluster and the cD galaxy were well subtracted, confirming the goodness of the spatial fit: notice the presence of a diffuse emission in the eastern part and of several discrete sources, numbered following Table 3.

**Fig.5**

Panel a): Temperature profile (in *keV*) as a function of the distance from the cluster center (in *arcmin*). The square (with dashed error bars) at the conventional distance of 0 *arcmin* represents the value for the dominant galaxy. In the last bin it is reported the value derived eliminating the contamination from the poor cluster (circle with solid error bars) and also the value for the whole bin (triangle with dashed error bars). For clarity of representation the triangle has been slightly shifted with respect to the x–coordinate. The dotted line represents the value of 3.25 $keV$, which is the mean of the first five bins.

Panel b): Abundance profile as a function of the distance from the cluster center (in $arcmin$); the cosmic abundance is $= 1$ in this scale. The symbols have the same meaning as in Panel a): notice that the circle in the last bin has no error bars because the abundance was fixed to the average value of 0.32, derived from the first five bins.

**Fig.6**

Relative behaviour of various mass components: solid lines and filled circles correspond to quantities derived with $H_o = 100\ km\ s^{-1}\ Mpc^{-1}$, while dotted lines and open circles refer to values obtained with $H_o = 50\ km\ s^{-1}\ Mpc^{-1}$. The vertical dashed line at 18 $arcmin$ (corresponding to $\sim 0.75\ h^{-1}\ Mpc$) indicates the radius beyond which the values have been obtained by extrapolating to larger radii the radial distribution of the gas density derived from the inner regions.

Panel a): Ratio between hot gas mass and stellar mass.

Panel b): Ratio between hot gas mass and dark matter component (computed as $M_{tot} - M_{gas} - M_{star}$).

Panel c): Ratio between dark and stellar matter.

Panel d): Ratio between total dynamical mass and stellar matter; note that in this panel the curves for the two $H_o$ values coincide.